\documentclass[%
 reprint,
 amsmath,amssymb,
 aps,prd,unsortedaddress,superscriptaddress,
 floatfix,caption,subcaption,
]{revtex4-2}
\usepackage{amsmath,amstext,amssymb}
\usepackage{float}
\usepackage[T1]{fontenc}
\usepackage{apjfonts} 
\usepackage{subfigure}
\usepackage{xcolor}
\usepackage{soul}
\usepackage{ulem}
\usepackage{comment}
\usepackage{graphicx}
\setstcolor{red}
\usepackage{lineno}
\usepackage{appendix}
\usepackage[artemisia]{textgreek}
\usepackage{orcidlink}
\usepackage{hyperref}
\graphicspath{{figures/}}

\newcommand{\appropto}{\mathrel{\vcenter{
  \offinterlineskip\halign{\hfil$##$\cr
    \propto\cr\noalign{\kern2pt}\sim\cr\noalign{\kern-2pt}}}}}

\begin{document}
\title{
Rapid Hierarchical Inference of Neutron Star Equation of State from multiple Gravitational Wave Observations of Binary Neutron Star Coalescences
}
\author{Anarya Ray\,\orcidlink{0000-0002-7322-4748
}}\email{anarya@uwm.edu}\affiliation{University of Wisconsin-Milwaukee, Milwaukee, WI 53201, USA}
\author{Michael Camilo}
\affiliation{Montclair State University, 1 Normal Ave, Montclair, NJ 07043 }
\author{Jolien Creighton\,\orcidlink{0000-0003-3600-2406}}
\affiliation{University of Wisconsin-Milwaukee, Milwaukee, WI 53201, USA}
\author{Shaon Ghosh\,\orcidlink{0000-0001-9901-6253}}
\affiliation{Montclair State University, 1 Normal Ave, Montclair, NJ 07043 }
\author{Soichiro Morisaki\,\orcidlink{0000-0002-8445-6747}}
\affiliation{Institute for Cosmic Ray Research, The University of Tokyo, 5-1-5 Kashiwanoha, Kashiwa, Chiba 277-8582, Japan}
\affiliation{University of Wisconsin-Milwaukee, Milwaukee, WI 53201, USA}

\begin{abstract}
     Bayesian hierarchical inference of phenomenological parameterized neutron star equations of state (EoS) from multiple gravitational wave observations of binary neutron star mergers is of fundamental importance in improving our understanding of neutron star structure, the general properties of matter at supra nuclear densities and the strong nuclear force. However, such an analysis is computationally costly as it is unable to re-use single-event EoS agnostic parameter estimation runs that are carried out regardless for generating gravitational wave transient catalogs. With the number of events expected to be observable during the 4th observing run (O4) of LIGO/Virgo/KAGRA, this problem can only be expected to worsen. We develop a novel and robust algorithm for rapid and computationally cheap hierarchical inference of parameterized EoSs from gravitational wave data which re-uses single event EoS agnostic parameter estimation samples to significantly reduce computational cost. We efficiently include a priori knowledge of neutron star physics as Bayesian priors on the EoS parameters. The high speed and low computational cost of our method allow for efficient re-computation of EoS inference every time a new binary neutron star event is discovered or whenever new observations and theoretical discoveries change the prior on EoS parameters. We test our method on both real and simulated gravitational wave data to demonstrate its accuracy. We show that our computationally cheap method produces EoS constraints that are completely consistent with existing analysis for real data, the chosen fiducial EoS for simulated data. Armed with our fast analysis scheme, we also study the variability of EoS constraints with binary neutron star properties for sets of simulated events drawn in different signal-to-noise ratio and mass ranges.

\end{abstract}

\maketitle
\section{Introduction}
\label{intro}

 The interior of a neutron star (NS) is one of the most extreme environments known to exist in the observable universe. The hydrostatic equilibrium established by the NSs own gravity acting against the degeneracy pressure of neutrons in its core is described by the Tolman-Oppenheimer-Volkoff (TOV) equations~\cite{TOV}. In order to solve the TOV equations, we need an equation of state (EoS) that relates the neutron degeneracy pressure inside the NSs core to its density at zero temperature. Due to the rapid cooling of NSs via neutrino emission and hence their fast \textbeta-equilibration, finite temperature and dissipative corrections to the EoS are negligible, allowing for NS matter to be modeled as a perfect fluid~\cite{cool}. The EoS of cold matter at supra nuclear densities is thought to be universal and has a one-to-one correspondence with the mass radius relationship of NSs~\cite{mr-pe}. Hence, knowledge of the cold matter nuclear EoS is of extreme importance in understanding not only the structure of NSs, but also the general properties of matter at such densities. However, limitations in our understanding of the strong nuclear force restrict our knowledge of the NS EoS to a number of approximate models. Quantitatively, the pressure density and the implied mass radius relationship for an EoS model $\mathcal{E}$ becomes:
\begin{eqnarray}
    p &=& p_{\mathcal{E}}(\rho) \iff r = r_{\mathcal{E}}(m) \label{eos-pe}
\end{eqnarray} 
where $p,\rho$ are the pressure and rest mass density at some point in the NS core, and $m,r$ are the mass and radius of the NS.

The tabulated set of equation of state models constitutes a discrete collection of models based on different theoretical descriptions of nuclear theories (for a recent list of tabulated EoS models, see~\cite{mr-review}), allowing us to compare one model against the others. However, this does not give us the ability to fully survey the pressure-density parameter space of matter at extreme densities. Parameterized EoS models such as the spectral decomposition of the adiabatic index in powers of logarithmic pressure~\cite{spectral} and the piecewise polytropic parameterization~\cite{piecewise}, are more flexible in the sense that a chosen range of the parameters correspond to a continuum in the pressure density space, which can be constrained empirically, given data. However, the data required for inferring the EoS are challenging to acquire from controlled experiments. The high density of NS cores, orders of magnitude higher than the nuclear saturation density~\cite{mr-review}, is impossible to re-create in a terrestrial laboratory. This has resulted in the NS EoS remaining largely unconstrained for a long time~\cite{mr-review}.

Observations of systems involving neutron stars often enable physicists to extract information regarding the NS EoS from measurements of the mass and/or radii of the NSs in such systems~\cite{mr-review}. The one-to-one correspondence between the NS mass-radius relationship and the EoS pressure-density relationship translates simultaneous mass-radius measurement into a measurement of the NS EoS itself. Observations of electromagnetic (EM) signals from such astrophysical systems, such as X-ray pulsar observations~\cite{EM-Xray1,EM-Xray2,EM-Xray3,EM-Xray4}, thermonuclear bursts and quiescent low mass X-ray binaries~\cite{thnc-qxlmb} etc, have been used to simultaneously measure the mass and radii of NSs and hence the NS EoS. However, these analyses are model-dependent and can be subject to the systematics of the individual EM emission models~\cite{mr-review}. Mass measurements of NSs through pulsar timing experiments~\cite{PT-1}, have also been informative about the NS EoSs using the fact that an EoS model implies a particular maximum mass of NSs. However, such precise mass measurements are possible only for pulsars in binary systems which comprise only 10 percent of the galactic pulsar population~\cite{PT-limit1,PT-limit2}. Thus, observations of EM signals from astrophysical NS systems have provided us with a natural laboratory for exploring the properties of cold nuclear matter at extreme densities, albeit in a rather limited notion. For a review of EoS measurements through electromagnetic signals from astrophysical NS systems see~\cite{mr-review}. 

Observations of gravitational waves (GWs) from merging binary neutron stars (BNS)s at ground based GW observatories such as LIGO~\cite{LVK1} and Virgo~\cite{LVK2} has resulted in significant advances in the field of NS EoS measurement. The GW waveform, in addition to being sensitive to the component NS masses,  carries an imprint of finite size effects in the evolution of BNS systems. Thus analyzing GW data from ground based observatories allows for the simultaneous measurements of NS mass and tidal deformability (akin to the radius), that are independent of EM emission model systematics. Hence GW based NS  mass-radii measurements yield EoS measurements that are free of the limitations of EM signal based EoS measurements. The universality of cold nuclear matter EoS can be exploited to combine GW observations from multiple BNS events together, or along with other astrophysical and terrestrial observations, to yield joint measurements of the NS EoS. GW data from the two merging BNSs observed to date, GW170817~\cite{170817disc} and GW190425~\cite{190425}, analyzed together with EM counterpart observations of GW170817~\cite{170817Adisc1,170817Adisc2,ATdisc}, other EM based astrophysical observations, terrestrial measurements and theoretical considerations have yielded joint EoS constraints that have greatly improved our understanding of dense nuclear matter~\cite{multieos1917}. With future GW observations from more BNSs expected to further improve these measurements, a robust and efficient scheme EoS inference, jointly from the GW observations of multiple BNS mergers, needs to be developed and/or perfected before the next observing run, O4, of the LIGO/Virgo/KAGRA~\cite{LVK3} (LVK) detector network begins.

Bayesian hierarchical inference from GW data~\cite{thranetalbot2019} is a statistical framework that has been used to accurately measure the EoS from multiple GW observations. However, numerical implementations of such an analysis incur a heavy computational cost, one that grows with the number of events analyzed. Thus, such implementations are likely to be problematic given the number of events expected to be observable during O4, potentially requiring several weeks of computation time. Several techniques~\cite{gmm,interpolation,interpolation2,interpolation3,gwxteme} have been developed to circumvent this problem that achieve their speed-up gains by re-using information from single event EoS agnostic parameter estimation (PE) runs that are carried out regardless for GW transient catalogs. Among them, the likelihood approximation algorithm \textsc{GWXtreme}~\cite{gwxtreme-doc} has been shown to perform accurate Bayesian model comparison on the tabulated set of known EoS models with a latency of a few minutes per event~\cite{gwxteme}. However, being a model selection scheme, the current version of \textsc{GWXtreme} cannot compute empirical credible intervals on the EoS pressure-density space from the data using parameterized EoS models. It can only comment on how some discrete lines on that space, one corresponding to each known EoS model, compare against each other.

In this work, we develop an algorithm based on \textsc{GWXtreme} to hierarchically infer parameterized EoSs from multiple GW observations. We incorporate our physical knowledge of NS physics such as the requirements of causality, thermal stability, and observational consistency of the NS maximum mass as Bayesian priors on the EoS parameters and compute their posterior distributions using our generalized version of \textsc{GWXtreme}, with a run time of 20 hours to a day for order 10 events. The posterior samples of EoS parameters then translate naturally into empirically measured confidence intervals in the EoS pressure-density plane. Since the \textsc{GWXtreme} method is based on an approximation scheme~\cite{gwxteme} rather than distribution of computation over large number of computational resources, we do not need GPU parallelization to achieve our results, making this method easier to implement on most machines. Our analysis needs no prior assumption about the fiducial population of BNSs.

Following previous works~\cite{interpolation,gmm} we test our algorithm on a set of simulated events drawn from the galactic BNS population and demonstrate its high accuracy. In addition, we test our algorithm on data from the real events, GW170817 and GW190425, to show that the results produced by our algorithm are fully consistent with the existing un-approximated Bayesian PE results, despite being orders of magnitude faster. Furthermore, armed with this fast and cheap algorithm we study the variability of EoS constraints with component NS masses, the signal-to-noise ratio (SNR) of events analyzed and the number of events analyzed by drawing additional simulated events in narrow SNR bins and over a wide range of BNS total masses. We note that this work is proof of concept, where-in we show that \textsc{GWXtreme}'s likelihood approximation scheme can be generalized to perform fast, cheap and accurate hierarchical inference of parameterized NS EoS models from multiple GW observations, making \textsc{GWXtreme} a strong candidate for hierarchical EoS inference in O4. We leave further improvements to our algorithm for potentially more generalized EoS inference as part of a future work, while outlining the blueprints of such generalizations.

This paper is organized as follows. In Sec.~\ref{methods}, we describe our method of EoS inference in detail. First, we review the Bayesian hierarchical framework for EoS inference from GW data in Sec.~\ref{bh} and the problematic computational cost of its implementation. Then, in Sec.~\ref{approx} we describe our algorithm, its approximations and how it resolves the aforementioned problem. Next, Sec.~\ref{eos-prior} summarizes the EoS parameterization and the priors on the EoS parameters that we have chosen for studying and testing our algorithm. We wrap up the discussion of methods in Sec.~\ref{waveform} with a note on the compatibility of our method with various GW waveform models and the associated systematics. In Sec.~\ref{data} we describe the details of the data on which we test our algorithm, which include both real and simulated events. In Sec.~\ref{results}, we display the results of our study and discuss their implications. In Sec.~\ref{conclusion} we conclude with a summary of our method, its virtues and its potential for performing efficient hierarchical EoS inference in O4 given the results of our study. We also outline schemes for potential improvements to our algorithm which are left as upcoming explorations.

\section{Methods}

\label{methods}
\subsection{Bayesian Hierarchical Inference of Parameterized EoS}
\label{bh}
In this section we review the framework of Bayesian hierarchical inference using GW data in the context of EoS inference. For a review of Bayesian inference from GW data see~\cite{thranetalbot2019}. Time series data from a GW detector, $d(t)$ can be thought to comprise random noise $n(t)$, with a modeled distribution, and possibly a signal $h(t,\vec{\theta})$ characterized by the parameter $\theta$ as dictated by the assumed signal model:
\begin{equation}
    d(t) = h(t,\vec{\theta})+n(t)\label{data0}
\end{equation}
The parameters $\vec{\theta}$ consist of both EoS sensitive observables (masses and  tidal deformability) and other observables such as distance, sky position, etc., to which the EoS is not sensitive. The finite size effects of NSs manifest themselves in the data through the dependence of the waveform model on the component NS masses $m_i$ and the tidal deformability of the NSs, $\lambda_i$:
\begin{eqnarray}
    \vec{\theta} &=& \{m_1,m_2,\lambda_1,\lambda_2,\vec{\theta}_{\text{ne}}\}\\
    \lambda_i &=&\frac{2}{3G} k_2(m_i)r_i^5\label{lambda}
\end{eqnarray}
where $r_i$, are radii of the component NSs, $k_2$ the tidal love number~\cite{love-number}, and $\vec{\theta}_{\text{ne}}$ are other non EoS sensitive parameters that characterize the GW waveform. Given a noise model which specifies the probability distribution of the random noise time series $n(t)$, one can compute the likelihood of the observed time-series data as a function of the parameters characterizing the waveform model: $\mathcal{L}(d|\vec{\theta})$. Since the EoS only constrains the EoS sensitive observables, we can marginalize the likelihood over the remaining parameters $\vec{\theta}_{\text{ne}}$ using uninformative priors $p(\vec{\theta}_{\text{ne}})$ and construct the ingredients of our EoS inference analysis from that marginalized likelihood:
\begin{equation}
    \mathcal{L}(d|m_1,m_2,\lambda_1,\lambda_2) = \int \mathcal{L}(d|m_1,m_2,\lambda_1,\lambda_2,\vec{\theta}_{\text{ne}})p(\vec{\theta}_{\text{ne}})\,d\vec{\theta}_{\text{ne}}\label{likelihood0}
\end{equation}
An EoS model, $\mathcal{E}$ implies a deterministic relationship between $m_i$ and $\lambda_i$, which imposes a delta-function prior on these quantities through Eq.~\eqref{eos-pe}: 
\begin{equation}
    p(\lambda_1,\lambda_2|m_1,m_2,{\mathcal{E}})=\delta(\lambda_1-\lambda_{\mathcal{E}}(m_1))\delta(\lambda_2-\lambda_{\mathcal{E}}(m_2))\label{prior-ms}
\end{equation}
where, the tidal parameter $\lambda$ as a function of mass is obtained by substituting Eq.~\eqref{eos-pe} into Eq.~\eqref{lambda}

Using the likelihood of GW data given EoS sensitive BNS parameters and an EoS model that imposes a prior on those parameters, one can construct the Bayesian evidence of the EoS model by marginalizing the likelihood over the prior:
\begin{widetext}
\begin{equation}
    \mathcal{Z}(\mathcal{E}|d) = \int \mathcal{L}(d|m_1,m_2,\lambda_1,\lambda_2)p(\lambda_1,\lambda_2|m_1,m_2,\mathcal{E})p(m_1,m_2)\,dm_1\,dm_2\,d\lambda_1\,d\lambda_2\label{evidence}
\end{equation}
\end{widetext}
where $p(m_1,m_2)$ is an uninformative prior on the component masses. The overall evidence of an EoS model given multiple observations $\{d\}=\{d_1,d_2,...\}$ can be obtained by multiplying the individual evidences, yielding the joint evidence: $\mathcal{Z}(\mathcal{E}|\{d\})=\prod_{i}\mathcal{Z}(\mathcal{E}|d_{i})$. The ratio of the joint evidence for two different EoS models $\mathcal{E}_{1}$ and $\mathcal{E}_2$ yields the Bayes factor which can be compared against unity to perform model selection:
\begin{equation}
    BF^{\mathcal{E}_1}_{\mathcal{E}_2}(\{d\}) = \frac{\mathcal{Z}(\mathcal{E}_1|\{d\})}{\mathcal{Z}(\mathcal{E}_2|\{d\})}
\end{equation}
This framework was used to perform model comparison for the set of known tabulated EoSs for the two BNS events observed by LVK to date: GW170817~\cite{170817MS}, and GW190425. However, model selection can only compare known EoS models with each other. While such an analysis sheds light on the physics of the NS EoS since each model has its own set of assumptions about said physics, empirical constraints on the pressure density space in the form of continuous credible intervals inferred from data are beyond its scope. Phenomenologically parameterized EoSs such as the spectral parameterization and the piecewise polytropic parameterization are free of this limitation.

Such an EoS model $\mathcal{E}$, characterized by the parameters $\vec{\gamma}$, would imply a pressure-density and the corresponding mass-tidal deformability relation:
\begin{eqnarray}
    p &=& p(\rho,\vec{\gamma})\label{pe-param}\implies
    \lambda =\lambda(m,\vec{\gamma})\label{ml-param}
\end{eqnarray}
The deterministic relation Eq.~\eqref{ml-param} would then impose a prior on $\lambda_i$ conditional on $m_i$ in the form of
\begin{eqnarray}
    p(\lambda'_1,\lambda'_2|m_1,m_2,\vec{\gamma})&=&\delta(\lambda'_1-\lambda(m_1,\vec{\gamma}))\delta(\lambda'_2-\lambda(m_2,\vec{\gamma}))\label{prior-pe}
\end{eqnarray}
Replacing the EoS sensitive prior in Eq.~\eqref{evidence} with Eq.~\eqref{prior-pe} would allow for the reinterpretation of the left hand side of Eq.~\eqref{evidence} as the marginalized hierarchical likelihood $\mathcal{L}_h$ of GW data $d$, given the EoS parameters $\vec{\gamma}$ (viewed as hyper-parameters):
\begin{widetext}
\begin{equation}
    \mathcal{L}_{h}(d|\vec{\gamma}) =  \int \mathcal{L}(d|m_1,m_2,\lambda_1,\lambda_2)p(\lambda_1,\lambda_2|m_1,m_2,\vec{\gamma})p(m_1,m_2)\,dm_1\,dm_2\,d\lambda_1\,d\lambda_2\label{likelihood}
\end{equation}
\end{widetext}
The universality of the cold matter nuclear EoS then implies that the EoS hyper-parameters $\vec{\gamma}$ are also universal and will have the same value for all NS systems. This can be exploited to construct the joint ``quasi'' likelihood of GW data from multiple events, $\{d\}$ given EoS hyper-parameters by multiplying the individual event hierarchical likelihoods. One can then use Bayes Theorem to convert that joint likelihood into a posterior distribution of EoS hyper-parameters given GW data from multiple events, by imposing a prior on those hyper-parameters based on our knowledge of NS physics:
\begin{equation}
    p(\vec{\gamma}|\{d\}) \propto  p(\vec{\gamma},I)\prod_{i=1}^N \mathcal{L}_{h}(d_i|\vec{\gamma})\label{posterior} 
\end{equation}
where $p(\vec{\gamma},I)$ encodes our a priori knowledge of NS physics by vanishing at values of $\vec{\gamma}$ for which the EoS becomes unphysical (examples of unphysicallity include acausal sound speed, thermal instability etc.). The abstract symbol $I$ represents our understanding of NS physics, to which our priors on the EoS hyper-parameters are conditional. This posterior distribution of the EoS hyper-parameters can be used to compute their Bayesian credible intervals which translate naturally into a credible interval on the EoS pressure-density space.

To compute credible intervals on the EoS hyper-parameters, one can in principle use standard Bayesian inference engines (such as direct quadrature or MCMC) on Eq.~\eqref{posterior} directly. However, the product of integrals in Eq.~\eqref{posterior} is difficult to deal with due to the presence of the delta functions in the integrands. The delta function priors imposed by the EoS model prevents us from approximating each integral in the product as a Monte Carlo sum over posterior samples yielded by single event EoS agnostic PE runs, that are carried out for generating GW transient catalogs. Unable to re-use information from single event PE runs, the alternative we are left with is to essentially redo the PE of BNS waveform parameters with delta function priors imposed on a subset of those parameters. This boils down to simultaneous inference of EoS hyper-parameters and the event specific waveform parameters, for all events. Numerical implementations of such PE would then require a large number of costly  GW template waveform evaluations per event. This would lead to a rapidly increasing computational cost of hierarchical EoS inference with the number of events analyzed, potentially requiring several weeks of computation time per analysis in O4.

It would be much more efficient if an algorithm can be developed to numerically compute the marginalized hierarchical likelihood as a fast evaluating function of its arguments, by somehow re-using information from single event EoS agnostic PE runs, which are carried out regardless, for GW transient catalogs. In the next section, we describe how to achieve this using the \textsc{GWXtreme} likelihood approximation scheme and the algorithm we developed based on it.

\subsection{Likelihood Approximation Scheme}
\label{approx}
To evaluate the hierarchical likelihood of GW data given EoS parameters, we perform the integral in Eq.~\eqref{likelihood} numerically, after approximating the integrand in a way that drastically reduces computational cost and latency. We base our approximation on~\cite{gwxteme}, where reparameterization of the EoS sensitive observables to reduce the dimensionality of the integrand and KDEs to approximate the lower dimensional (marginalized) integrand were implemented, for rapid and computationally cheap computation of EoS evidences, i.e., Eq.~\eqref{evidence}. Since the integral in Eq.~\eqref{likelihood} is essentially the same as that in Eq.~\eqref{evidence} with the only difference being the parameterization of the EoS sensitive prior, similar approximations as in~\cite{gwxteme} can be used for the fast and cheap evaluation of our hierarchical marginalized likelihood.

To reduce the dimensionality of the integral in Eq.~\eqref{likelihood} and evaluate it numerically, we first note that by Bayes theorem, the single event likelihood given EoS sensitive BNS parameters, multiplied by EoS uninformative priors on those parameters, is proportional to the posterior distribution of those parameters given GW data: $p(m_1,m_2,\lambda_1,\lambda_2|d) \propto \mathcal{L}(d|m_1,m_2,\lambda_1,\lambda_2,\vec{\theta}_{\text{ne}} ) p(m_1,m_2,\lambda_1,\lambda_2)$. Note that this posterior has already been sampled during the single event EoS agnostic PE run for some choice of the uninformative priors: $p(m_1,m_2,\lambda_1,\lambda_2)=p_{\text{PE}}(m_1,m_2,\lambda_1,\lambda_2)$. This can be used to rewrite the integral in Eq.~\eqref{likelihood} in terms of the single-event posterior distribution of BNS parameters $p(m_1,m_2,\lambda_1,\lambda_2|d)$:
\begin{widetext}
\begin{equation}
    \mathcal{L}_h(d|\vec{\gamma}) = \int \frac{p(m_1,m_2,\lambda_1,\lambda_2|d)}{p_{\text{PE}}(m_1,m_2,\lambda_1,\lambda_2)}p(\lambda_1,\lambda_2|m_1,m_2,\vec{\gamma})p(m_1,m_2)\,dm_1\,dm_2\,d\lambda_1\,d\lambda_2\label{likelihood2}
\end{equation}
\end{widetext}
As we shall demonstrate shortly, the single-event posterior distribution of BNS parameters given GW data can be accurately approximated as a fast evaluable numerical function, from its stochastic samples which are already generated and written to disk during the creation of GW transient catalogs. Before implementing such an approximation, we first show  that the dimensionality of the integral can be now be reduced by reparameterizing the EoS sensitive observables. We re-parameterize $(m_1,m_2,\lambda_1,\lambda_2)$ into the chirp mass, mass ratio, and tidal parameters, $\mathcal{M}(m_1,m_2),q(m_1,m_2),\tilde{\Lambda}(\lambda_1,\lambda_2,m_1,m_2),\delta\tilde{\Lambda}(\lambda_1,\lambda_2,m_1,m_2)$, which are defined as follows:
\begin{widetext}
\begin{eqnarray}
    \mathcal{M} &=& \frac{(m_1m_2)^{3/5}}{(m_1+m_2)^{1/5}}\\
    q &=& m_2 / m_1 \\
    \tilde{\Lambda} &=& \frac{8}{13}\big[(1+7\eta -31\eta^2)(\Lambda_1+\Lambda_2)+\sqrt{1-4\eta}(1+9\eta-11\eta^2)(\Lambda_1-\Lambda_2)\big]\label{Lt}\\
    \delta\tilde{\Lambda} &=&\frac{1}{2}\bigg[ \sqrt{1-4\eta}\left(1-\frac{13272}{1319}\eta+\frac{8944}{1319}\eta^2\right)(\Lambda_1+\Lambda_2)+\left(1-\frac{15910}{1319} \eta +\frac{32850}{1319}\eta^2 +\frac{3380}{1319}\eta^3\right)(\Lambda_1-\Lambda_2)\bigg]\label{dLt}
\end{eqnarray}
where $\eta=m_1m_2/(m_1+m_2)^2$ is the symmetric mass ratio and $\Lambda_i=G\lambda_i[c^2/(Gm_i)]^5$ is the dimensionless tidal deformability. We note that the definitions Eq.~\eqref{Lt} and Eq.~\eqref{dLt} assume $m_1>m_2$.  Under this reparameterization, Eq.~\eqref{likelihood2} becomes:
\begin{equation}
    \mathcal{L}_h(d|\vec{\gamma}) \propto \int \frac{p(\mathcal{M},q,\tilde{\Lambda},\delta\tilde{\Lambda}|d)}{p_{\text{PE}}(\mathcal{M},q,\tilde{\Lambda},\delta\tilde{\Lambda})}p(\tilde{\Lambda},\delta\tilde{\Lambda}|\mathcal{M},q,\vec{\gamma})p(\mathcal{M},q)\,d\mathcal{M}\,dq\,d\tilde{\Lambda}\,d\delta\tilde{\Lambda}\label{likelihood3}
\end{equation}
where the Jacobians associated with the variable change cancel out and the reparameterized EoS sensitive priors are
\begin{equation}
   p(\tilde{\Lambda}',\delta\tilde{\Lambda}'|\mathcal{M},q,\vec{\gamma})= \delta(\tilde{\Lambda}'-\tilde{\Lambda}(\mathcal{M},q,\vec{\gamma}))\delta(\delta\tilde{\Lambda}'-\delta\tilde{\Lambda}(\mathcal{M},q,\vec{\gamma}))
\end{equation}
\end{widetext}
To further simplify Eq.~\eqref{likelihood3}, we can choose the uninformative priors in the denominator of the integrand to be uniform in the reparameterized tidal deformabilities: $p_{\text{PE}}(\mathcal{M},q,\tilde{\Lambda},\delta\tilde{\Lambda})\propto p(\mathcal{M},q)$. This choice is compatible with at least one standard GW waveform for BNSs: the \textsc{TaylorF2} Waveform model~\cite{TaylorF2}. We elaborate more on Generalizations to other waveforms and the associated systematics in Sec.~\ref{waveform}. Under this choice of priors, Eq.~\eqref{likelihood3} becomes
\begin{equation}
    \mathcal{L}_h(d|\vec{\gamma}) \propto \int p(\mathcal{M},q,\tilde{\Lambda},\delta\tilde{\Lambda}|d)p(\tilde{\Lambda},\delta\tilde{\Lambda}|\mathcal{M},q,\vec{\gamma})\,d\mathcal{M}\,dq\,d\tilde{\Lambda}\,d\delta\tilde{\Lambda}
    \label{likelihood31}
\end{equation}
The reason for this reparameterization is as follows. The chirp mass $\mathcal{M}$ is known to be extremely well measured, with its posterior distribution being sharply peaked about a number equal to the mean of the samples of $\mathcal{M}$ which are available from single event PE runs. Among the tidal parameters, $\tilde{\Lambda}$, which enters the GW waveform model at the 5th post-Newtonian order, has the dominant contribution as compared to $\delta\tilde{\Lambda}$, which enters the waveform at the 6th post-Newtonian order. As a result, the posterior distribution of BNS parameters is largely independent of $\delta\tilde{\Lambda}$. Under these considerations, the posterior distribution of BNS parameters given GW data can be approximated as $p(\mathcal{M},q,\tilde{\Lambda},\delta\tilde{\Lambda}|d)\approx p(q,\tilde{\Lambda}|d)\delta(\mathcal{M}-\bar{\mathcal{M}})$, where $\bar{\mathcal{M}}$ is the mean of the chirp mass samples obtained from EoS agnostic single event PE runs. Substituting these into Eq.~\eqref{likelihood3} allows us to use the delta functions and reduce the dimensionality of the integral to one, provided the marginalized posterior $p(q,\tilde{\Lambda}|d)$ can be evaluated, at least approximately, as a numerical function of its parameters.
\begin{equation}
    \mathcal{L}_h(d|\vec{\gamma}) \appropto \int p(q,\tilde{\Lambda}(\bar{\mathcal{M}},q,\vec{\gamma})|d)\,dq\label{likelihood4a}
\end{equation}

\begin{figure}[t!]
    \centering
    \subfigure[SNR=32]{\includegraphics[width=0.5\textwidth]{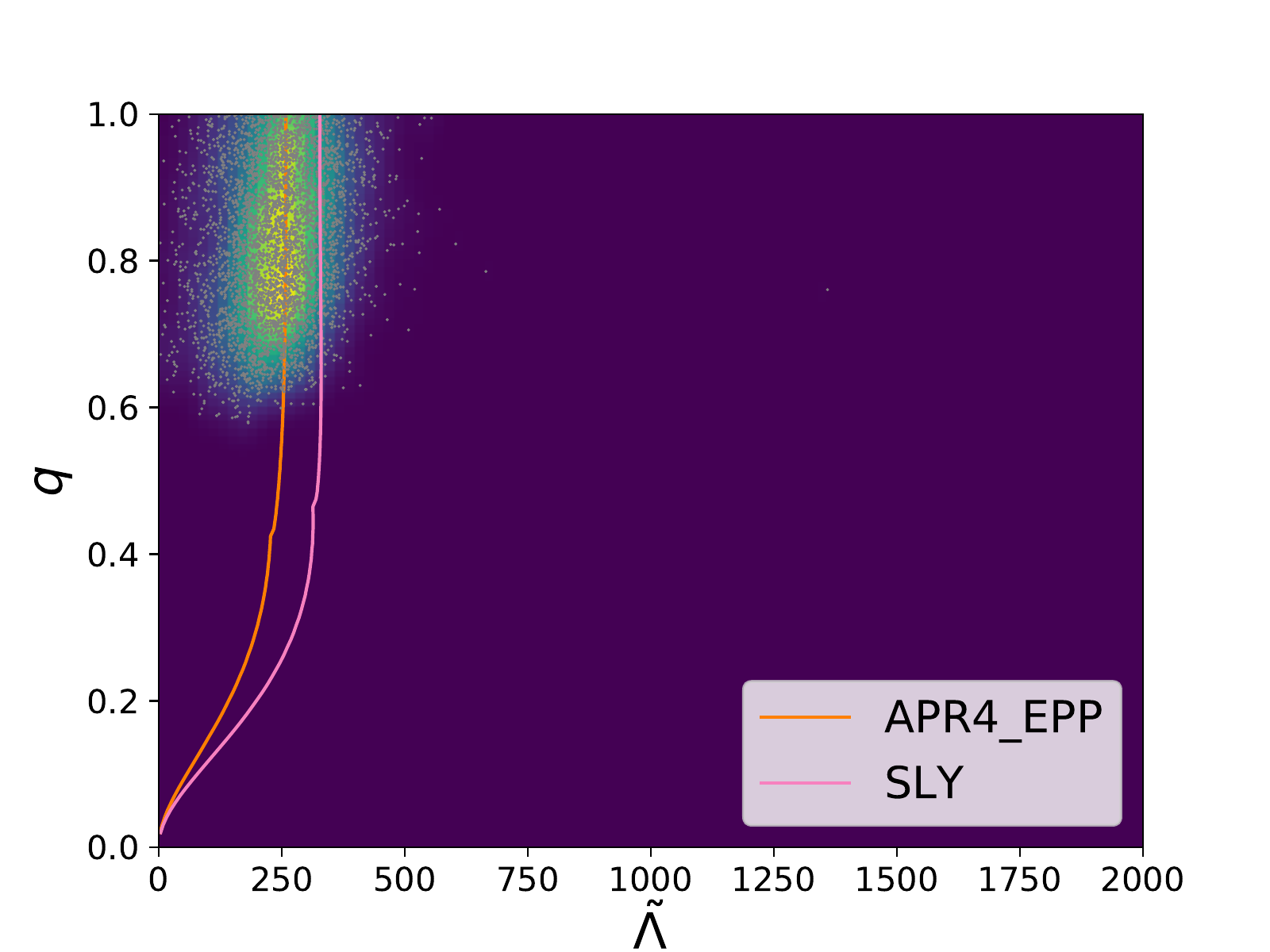}}
    \subfigure[SNR=11]{\includegraphics[width=0.5\textwidth]{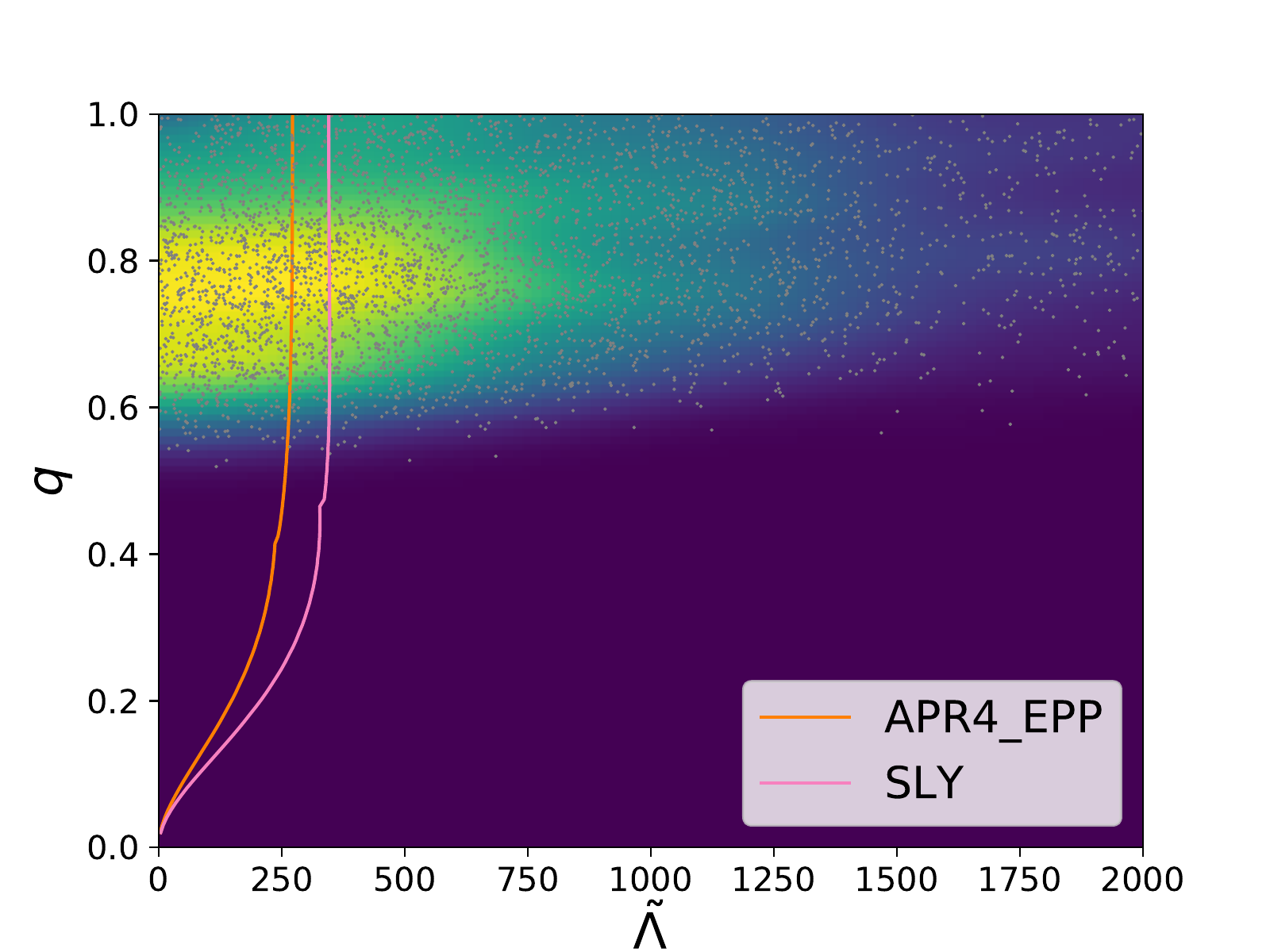}}
\caption{KDE visualization for two simulated events, with APR4\_EPP as the injected EoS, one at low (bottom) and the other at high (top) SNR. In both the figures we have plotted $K_i(q,\tilde{\Lambda})$ as a 2D density plots. We have overplotted the PE samples of $q,\tilde{\Lambda}$ as discrete gray points to demonstrate the KDE accurately approximates the posterior distribution from which the samples are drawn. The integral in Eq.~\eqref{posterior2} can be visualized as line integrals of these 2D density along the $\tilde{\Lambda}(q,\mathcal{M},\vec{\gamma})$ line for some particular value of $\vec{\gamma}$. Two such lines, one corresponding to the injected EoS APR4\_EPP and a different SLY are plotted as an example of the EoS lines along which the integrals are performed. }
    \label{kde-vis}
\end{figure}

To evaluate $p(q,\tilde{\Lambda}|d)$ as a numerical function of $(q,\tilde{\lambda})$, we approximate it from EoS agnostic single event PE samples of its parameters via Gaussian kernel density estimation (KDE), customized to be immune to edge effects. Traditional Gaussian KDE fits a Gaussian around each posterior sample and approximates the density of those samples as the sum of those Gaussians~\cite{KDE1,KDE2}. The covariance matrix of each of the Gaussians is approximated from the sample covariance matrix of the posterior samples. While the KDE evaluation speed is very fast, it is susceptible to edge effects in its traditional form, becoming inaccurate for distributions with sharp edges. Single event PE  runs implicitly assign the heavier NS as the primary mass $m_1$, resulting in a sharp edge on the distribution of mass ratio at $q=1$. To circumvent the incompatibility of this sharp edge with KDE, we add to the KDE probability distribution function (pdf) at each point, the value of the pdf at a point symmetric about the sharp edge, resulting in a bounded KDE, similar to~\cite{gwxteme}. This results in our KDE being free of edge effects and compatible with the sharp edge at $q=1$. The accuracy of this approximation is demonstrated in Fig.~\ref{kde-vis}. With this approximation of the marginalized posterior, the hierarchical likelihood for the $i$th GW event becomes
\begin{equation}
    \mathcal{L}_{h}(d_i|\vec{\gamma}) \appropto \int_0^1 K_i(q,\tilde{\Lambda}(\bar{\mathcal{M}}_i,q,\vec{\gamma}))\,dq\label{likelihood4}
\end{equation}
where $K_i(q,\tilde{\Lambda})$ is the KDE approximation of $p(q,\tilde{\Lambda}|d_i)$ obtained from the EoS agnostic single event posterior samples $\{(q,\tilde{\Lambda})_i\}$. Since $K_i$ is a fast evaluating function of its arguments, the definite integral with a finite range in Eq.~\eqref{likelihood4}  can be evaluated efficiently using numerical techniques such as the trapezoidal rule. Substituting Eq.~\eqref{likelihood4} into Eq.~\eqref{posterior} yields the approximate joint hierarchical posterior of EoS parameters given GW data from multiple BNS observations, the EoS parameterization model and the prior knowledge on NS matter physics, that is numerically evaluable almost instantaneously: 
\begin{equation}
    p(\vec{\gamma}|\{d\},I) \appropto p(\vec{\gamma},\textbf{I})\prod_{i=1}^N \int_{0}^1 K_i(q,\tilde{\Lambda}(\bar{\mathcal{M}}_i,q,\vec{\gamma}))\,dq \label{posterior2}
\end{equation}

The posterior distribution in Eq.~\eqref{posterior2} can be sampled stochastically to produce empirical constraints on the EoS pressure density relation. We use affine-invariant Markov Chain Monte Carlo (MCMC) ensemble sampling~\cite{emcee1}, as implemented in the the package \textsc{emcee}, with CPU parallelization inbuilt~\cite{emcee2}, to sample the posterior in Eq.~\eqref{posterior2}. We parallelize emcee on 50 CPU cores to sample the posterior in Eq.~\eqref{posterior2} within 20 hours to a day, for 10 BNS events. Using those posterior samples, for each value of the density $\rho$ in Eq.~\eqref{pe-param}, we evaluate $N_{\text{samples}}$ number of pressures $\{p(\rho,\vec{\gamma}_j)\}$, one corresponding to every posterior sample $\{\vec{\gamma}_j\}$ where, $j=1,2,...,N_{\text{samples}}$. We then compute the median, 5 and 95 percent quantiles from the set $\{p(\rho,\vec{\gamma}_j)\}$, which we plot against $\rho$ and interpret them as empirical constraints/credible intervals on the NS EoS\@. The posterior predictive distribution of the dimensionless tidal parameter at $1.4\,M_{\odot}$ can also be produced as a different representation of how well the EoS is constrained. These are obtained by computing the histograms of the values of the dimensionless tidal parameter at $1.4\,M_{\odot}$ $\Lambda_i=\Lambda(m=1.4\,M_{\odot},\vec{\gamma_i})$ corresponding to each posterior sample $\vec{\gamma_i}$ of the EoS parameters drawn form the joint posterior Eq.~\eqref{posterior2}. We produce these constraints by performing our analysis on GW data from both real and simulated events and demonstrate the accuracy and efficiency of our method.

Even though we do not simultaneously infer BNS population parameters, we note that our algorithm is generalizable to do so while retaining its low computational cost. We discuss the blueprints of such a calculation in the conclusion section and leave it as part of an upcoming work.
In the next section, we describe our choice of EoS parameterization and the priors imposed on the EoS parameters, which are based on existing knowledge of NS matter physics.

\subsection{EoS parameterization and priors}
\begin{figure}[!ht]
    \centering
    \includegraphics[width=0.5\textwidth]{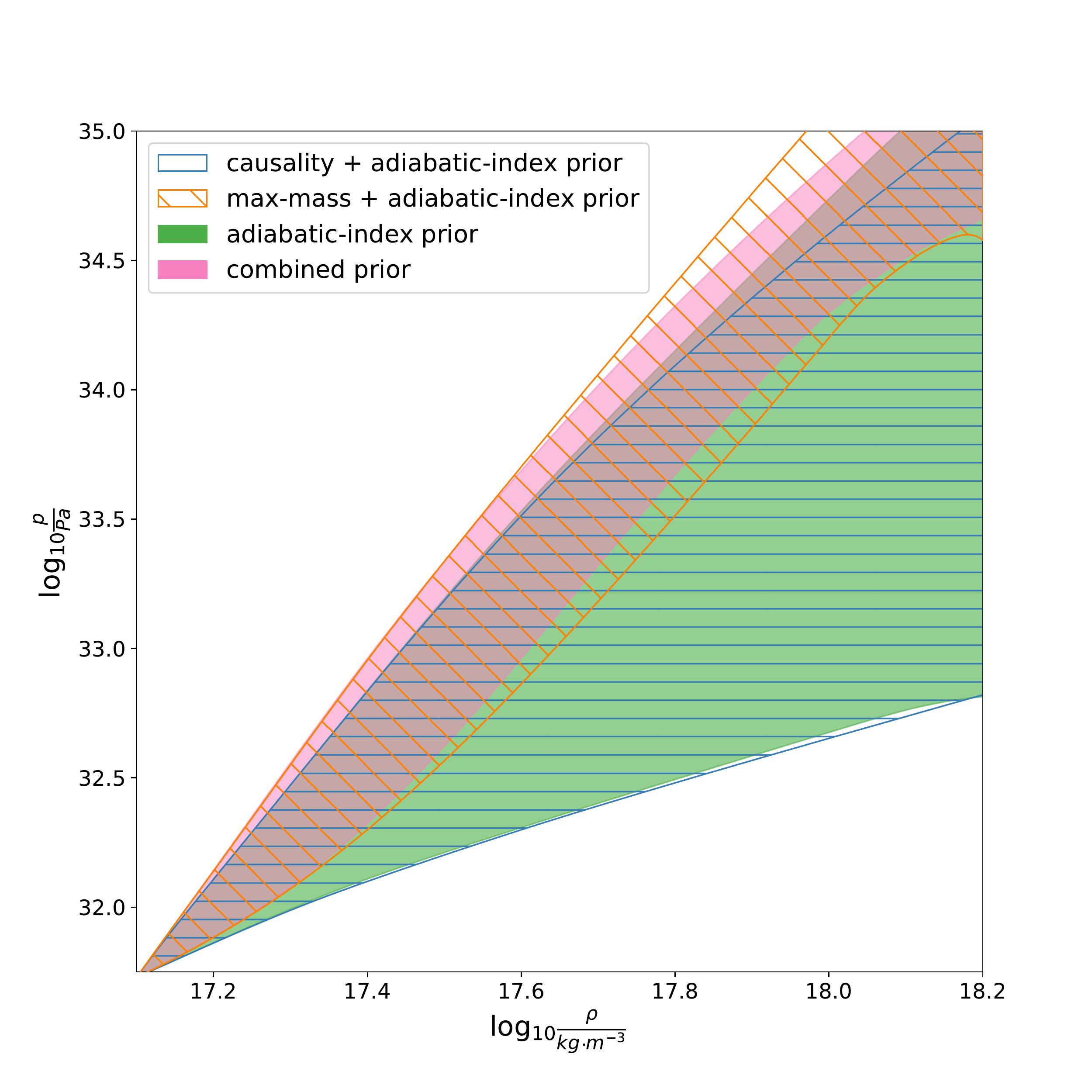}
    \caption{Visualization of the priors used on EoS parameters as bounds on the pressure density plane. We draw a large number of samples of the spectral parameters from the uniform distributions on the right hand side of Eq.~\eqref{prior}. From those samples, we only keep the ones that satisfy the physicality conditions imposed by the various component of the prior. We then calculate the pressure density curves for each of those samples and plot credible intervals on the pressure density plane that contain 90 percent of those curves.}
    \label{priors}
\end{figure}
\label{eos-prior}
Previous studies like~\cite{piecewiseonly} have shown the effectiveness of using phenomenologically parameterized EoSs such as the piecewise polytrope,  in measuring the EoS pressure-density relation empirically from GW data. Parameterizing the pressure density relation instead of the mass-tidal parameter relation and deriving the latter from the former makes the inclusion of a priori knowledge of NS physics into the analysis straightforward, in the form of Bayesian priors on the EoS parameters. However, the non-differentiability in the EoS at the joining point of two consecutive polytrope pieces leads to increases statistical errors in the EoS measurements inferred from GW data using the piecewise parameterizations, at those points~\cite{spectralpiecewise}. The spectral parameterization of the EoS which expands the adiabatic index in powers of logarithmic pressure, has been shown to be free of this deficiency~\cite{spectralpiecewise}. For this reason, we choose the four parameter spectral decomposition as our EoS parameterization for this study while noting that our analysis works for any parameterized model including the piecewise-polytrope, which we show in the appendix as a validation study. Under the spectral parameterization, the adiabatic index in terms of the pressure is
\begin{equation}
    \ln \Gamma(p,\vec{\gamma}) = \sum_{k=1}^{4}\gamma_k \left(\ln{\frac{p}{p_0}}\right)^k\label{spectral1}
\end{equation}
where $\Gamma(p)$, is the adiabatic index at pressure $p$ and $p_0$ is the minimum pressure above which the representation is valid. Eq.~\eqref{spectral1} can be used to directly compute pressure as a function of energy density $e$ using the thermodynamic relation
\begin{equation}
    \frac{de}{dp} =  \frac{e + p}{p\Gamma(p,\vec{\gamma})}
\end{equation}
which can then be used to find rest mass density as a function of pressure using  $\rho(p,\vec{\gamma})=[e(p,\vec{\gamma})+p]\exp[-h(p,\vec{\gamma})]$ where $dh/dp=1/[e(p,\vec{\gamma})+p]$. The pressure density relation can then be used to solve the TOV equation to yeild  $\tilde{\Lambda}=\tilde{\Lambda}(\mathcal{M},q,\vec{\gamma})$. We use the \textsc{LALSimulationNeutronStarEOS} module of the LIGO Algorithm Library package (\textsc{LALSuite})~\cite{lalsuite}, that implements TOV solving algorithms described in~\cite{mr-pe} and~\cite{TOVsol}, to solve the TOV equation, for each $\vec{\gamma}$.

As implemented in \textsc{LALSimulationNeutronStarEOS}, the minimum pressure $p_0$ is chosen to be  $5.37\times 10^{34}\,\text{dyn}\,\text{cm}^{-2}$. At pressures below this value, a different EoS, the SLY EoS model of~\cite{sly}, is used, which is stitched to the high density spectral EoS at $p=p_0$, as implemented in \textsc{LALSimulationNeutronStarEOS}, consistent with previous works like~\cite{spectralpiecewise,interpolation}. The range of $\vec{\gamma}$ for which the EoS is physical and stable can be determined using existing knowledge of NS physics. Following~\cite{spectralpiecewise} and~\cite{interpolation} we demand that for an EoS characterized by a particular value of the parameters $\vec{\gamma}$ to be physical and observationally consistent, it must be thermally stable, causal and result in a maximum NS mass that is larger than the mass of the most massive NS observed with precise mass measurement to date. The thermal stability requirement, which demands that $de/dp>0$, is satisfied by imposing $\Gamma(p,\vec{\gamma}) \in [0.6,4.5]$ for all $p\in (5.37\times10^{32},1.19\times 10^{38})\,\text{dyn}\,\text{cm}^{-2}$, in addition to the  uniform priors on the $\vec{\gamma}$: $\gamma_0 \in [0.2,2]$, $\gamma_1 \in [-1.6,1.7]$, $\gamma_2 \in [-0.6,0.6]$, $\gamma_3 \in [-0.02,0.02]$. The causality prior demands that the speed of sound in NS matter, $c_s=\sqrt{dp/de}$, up to and at the central pressure $p_{c,max}$ of the heaviest NS supported by the EoS, has to be less than the speed of light. We allow for a 10 percent buffer in the speed of sound inequality, to account for causal EoS models in the list of tabulated EoSs to be a fit by acausal spectral EoSs. Finally, we impose the maximum mass constraint by demanding that spectral parameters corresponding to physical EoSs must correspond to a maximum NS mass that is larger than that of the most massive NS observed, for which precise mass measurements are possible: $m_{\text{max}}(\vec{\gamma})>1.97\,M_{\odot}$~\cite{mmaxprior}. All of these priors are consistent with previous work on hierarchical inference of parameterized EoSs from GW data, for example~\cite{spectralpiecewise,piecewiseonly,interpolation}. We note that the recent mass measurements of PSR J0952-0607~\cite{heaviest} renders this choice of $1.97\,M_{\odot}$ for the heaviest observed NS mass outdated. However, since we validate our results for the real events by comparing with previous analyses like~\cite{170817Eos,190425} that were carried out before this recent observation, we choose $1.97\,M_{\odot}$ for the heaviest observed NS mass so as to be consistent with those works.

Combining all of these individual priors, our resultant prior on the EoS parameters is:
\begin{widetext}
\begin{equation}\label{prior}
    \renewcommand\arraystretch{0}
    p(\vec{\gamma},I) \propto \begin{cases}
    1 & \text{where}\; \begin{array}[t]{l} 0.2 \le \gamma_0 \le 2,\; -1.6 \le \gamma_1 \le 1.7,\; -0.6 \le \gamma_2 \le 0.6,\; -0.02 \le \gamma_3 \le 0.02, \\  0.6 \le \Gamma(p,\vec{\gamma}) \le 4.5,\;  c_{s}(p_{c,max},\vec{\gamma})<1.1c,\; \text{and}\;  m_{\text{max}}(\vec{\gamma})>1.97 M_{\odot} \end{array} \\
    0 & \text{otherwise}
    \end{cases}
\end{equation}
\end{widetext}

For a visualization of how these priors constrain the EoS pressure-density relation, we draw samples of the EoS parameters from the prior in Eq.~\eqref{prior} and compute pressure density credible intervals from those samples. We plot the intervals for 50000 combined prior samples in Fig.~\ref{priors}.

In the next section, we summarize the compatibility of our approximation, with various GW waveform models that might be used to perform the single-event PE.

\subsection{Waveform Systematics}
\label{waveform}
\begin{figure}[!hb]
    \centering
    \includegraphics[width=0.5\textwidth]{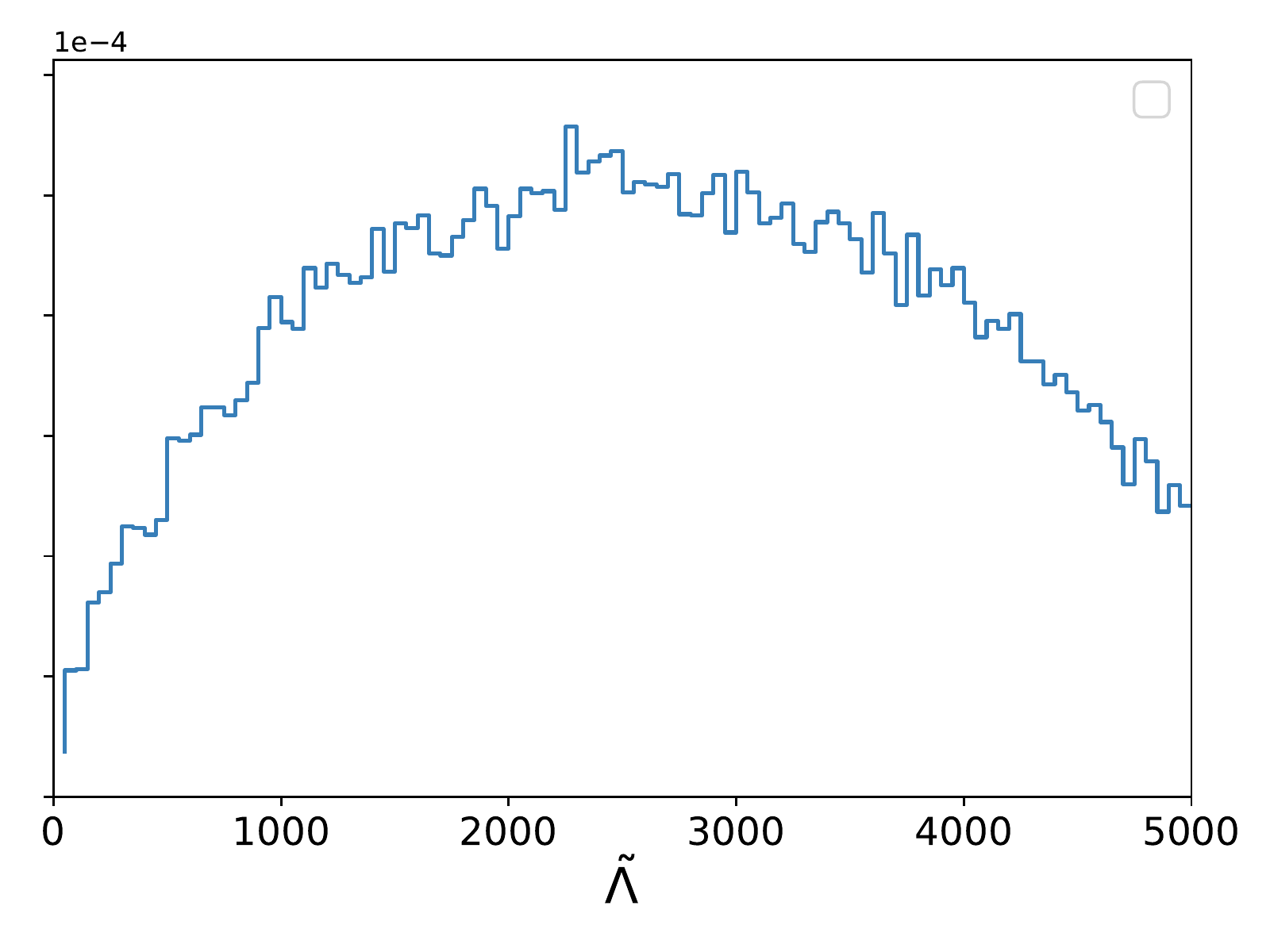}
    \caption{Visualization of uninformative priors on tidal parameters used in single event EoS agnostic PE runs which can be one or the other depending on the waveform model being used. The uniform in $\tilde{\Lambda}$ prior, compatible with \textsc{TaylorF2} and hence the current version of \textsc{GWXtreme}, would be a horizontal line in this plot. The uniform in positive $\Lambda_1$, $\Lambda_2$, which is compatible with the \textsc{PhenomPNRT} waveform families leads to a $\tilde{\Lambda}$ prior distribution that is shown in this plot and evidently has vanishing support at $\tilde{\Lambda}=0$. This can tend to blow up the integrand in Eq.~\eqref{likelihood3} as its denominator $p_{\text{PE}}(\mathcal{M},q,\tilde{\Lambda},\delta\tilde{\Lambda}$ will then be proportional to a quantity that vanishes a region inside the integration range. }
    \label{l1l2prior} 
\end{figure}

The compatibility of our approximation scheme with various GW waveform models manifests through the dependence of the choice of uninformative priors $p_{\text{PE}}(\mathcal{M},q,\tilde{\Lambda},\delta\tilde{\Lambda})$ used in the single-event PE runs on said waveform models. The equivalence of Eq.~\eqref{likelihood3} and Eq.~\eqref{likelihood31}  is conditional on whether the $p_{\text{PE}}(\mathcal{M},q,\tilde{\Lambda},\delta\tilde{\Lambda})$ is uniform in $\tilde{\Lambda}$. However, such a prior typically has support in regions of the $\tilde{\Lambda}$-$\delta\tilde{\Lambda}$ space that corresponds to negative values of the tidal parameters $\lambda_1$, $\lambda_2$, which can cause certain waveform generators for certain families of GW waveform models to fail. \textsc{PhenomNRT}~\cite{phenom1,phenom2,phenom3,phenom4}, is one such example. On the other hand, a uniform in $\Lambda_1$-$\Lambda_2$ prior, which is compatible to \textsc{PhenomNRT} waveforms, will imply a $p_{\text{PE}}(\mathcal{M},q,\tilde{\Lambda},\delta\tilde{\Lambda})$ that is not uniform in $\tilde{\Lambda}$ and has vanishing support at $\tilde{\Lambda}=0$, as can be seen in Fig.~\ref{l1l2prior}. This will tend to blow up the integrand in Eq.~\eqref{likelihood3} near $\tilde{\Lambda}=0$ which might result in unacceptable numerical errors.

The \textsc{TaylorF2} waveform~\cite{TaylorF2}, is immune to this issue since it is evaluated originally as a function of $(\tilde{\Lambda},\delta\tilde{\Lambda})$ and does not need a conversion to $\Lambda_1$-$\Lambda_2$ space. A uniform in $\tilde{\Lambda}$ prior for the single event PE runs is thus compatible with \textsc{TaylorF2}. Hence we choose the \textsc{TaylorF2} waveform model to perform our single event EoS agnostic PE runs for both real and simulated events.

Following~\cite{gwxteme,spectralpiecewise,piecewiseonly}, we truncate \textsc{TaylorF2}'s frequency domain waveform model at a stage of the binary's evolution, where the separation between the two NSs become comparable to the innermost stable circular orbit (ISCO), at which point the frequency of GW has the value $f_{\text{ISCO}}=c^3/[6^{2/3}\pi G(m_1+m_2)]$. While this choice of upper frequency cutoff can be unrealistic for EoSs that predict large NS radii, leading to merger happening before the separation of the NSs reaching ISCO, GW detectors are largely insensitive to such high frequencies, rendering the unphysicallity of our frequency cutoff irrelevant. It has been shown in previous studies, such as~\cite{fcutoff}, that varying the cutoff frequency has a negligible effect on the EoS Inference from GW data. 

We note that our algorithm can be made compatible with other waveforms such as \textsc{PhenomNRT} that require positive $\Lambda_1$ and $\Lambda_2$ by switching to a 3-dimensional KDE, as described in Sec.~\ref{conclusion}. We leave such a generalization as part of an upcoming work. In the next section, we describe in detail the simulation study we performed, the single event PE method used in those simulations and the real event data, on which we run our analysis to test its accuracy. 

\section{Data: Simulation Study and real events}
\label{data}
\begin{figure}[!ht]
   \subfigure[SNR $\in (23,25)$]{\includegraphics[width=0.49\textwidth]{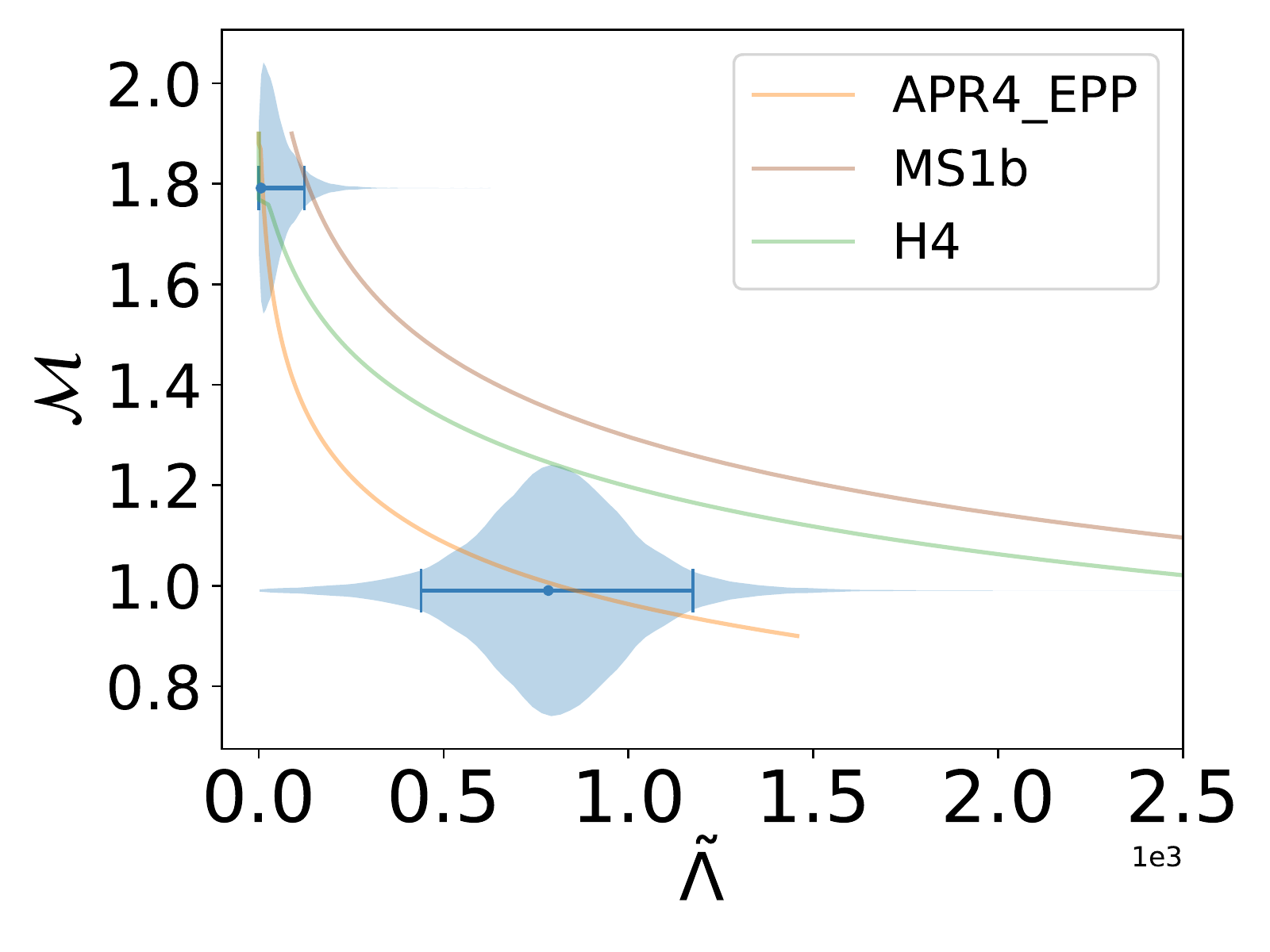}}
    ~
   \subfigure[SNR $\in (33,35)$]{\includegraphics[width=0.49\textwidth]{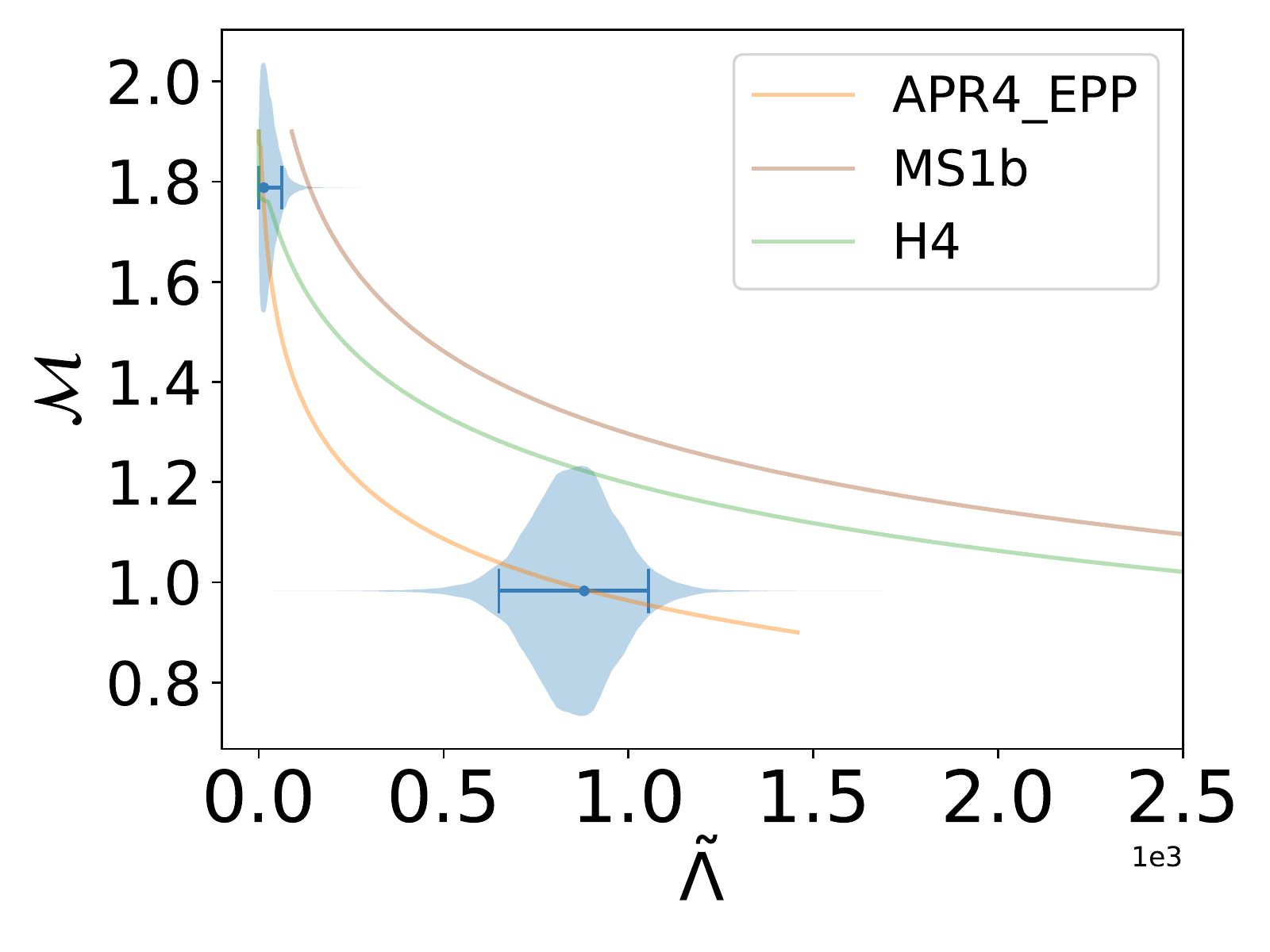}}
    \caption{Visualization of why high mass events are less informative on the EoS. We plot 90\% highest posterior density intervals of the $\tilde{\Lambda}$ posteriors for a low mass and a high mass simulated event drawn from the narrow SNR bins, top: (23,25) and bottom: (33,35). We also display the shape of the posterior in the form of violin plots. The solid lines are the $\mathcal{M}$-$\tilde{\Lambda}$ curves at $q=1$ for some of the known EoS models.}
    \label{lo-hi}
\end{figure}
We run our algorithm on the single event PE data for real events GW170817 and GW190425 as released publicly by the LVC~\cite{170817PE,190425PE}, both individually and jointly. We compare our results with the EoS constraints obtained from the full unapproximated PE run that infers the event-specific BNS parameters simultaneously with the EoS hyper-parameters. We show that the results from our approximate technique are in very good agreement with that of the full PE results, while being orders of magnitude faster. To test the accuracy of our algorithm in recovering the true EoS from data we run our analysis on simulated events which were chosen as follows.

We randomly draw BNS masses from the Galactic BNS population as inferred in~\cite{pop}.  Astrometric parameters were chosen in such a way that the sky positions and orientations of the coalescences are isotropic and distributed uniformly in co-moving volume within a luminosity distance range of 30\,Mpc to 200\,Mpc. This is consistent with previous work:~\cite{gmm,interpolation} except for the distance distribution which~\cite{gmm} chose to be uniform in distance instead. However, unlike the aforementioned works, we also set the dimensionless spin parameters to be zero for simplicity, since neutron stars are expected to spin down and loose rotational energy to magnetically driven plasma winds~\cite{spin1,spin2,spin3}. This choice is further justified by noting that galactic double neutron star systems have been observed to have very low spins (with dimensionless spin-parameter $\chi<0.05$)~\cite{pop2,spinpop}. We assign our fiducial/``true'' EoS to be APR4\_EPP~\cite{APR4EPP} and compute the tidal parameters using said EoS from the drawn masses. Then, using these parameters and the \textsc{TaylorF2} waveform model, we simulate a GW waveform and inject it into GW detector noise realizations corresponding to power spectral densities of the 4 LVK detectors at projected O4 sensitivity~\cite{O4projection}.

Among the events drawn, we choose the first $N_{\text{O4}}$ with a signal-to-noise ratio greater than or equal to 8 in at least one detector and perform PE on them to infer the EoS agnostic posteriors of single event BNS parameters given the simulated GW data. Here the  expected number of observable events in O4, $N_{\text{O4}}$, is calculated using Poisson statistics, the number of events discovered till date and the projected O4 sensitivity, as derived in appendix~\ref{poisson}. With 2 confident BNSs observed till date and the projected O4 sensitivity estimated in ref~\cite{O4projection}, we find the upper bound on $N_{\text{O4}}$ to be 16 with 90\% confidence. This effectively makes our EoS constraints obtained from analyzing the simulated events a forecast for how well we might be able to constrain the NS EoS from GW data alone, by the time O4 ends. 

To study the variability of EoS constraints we also draw equal mass simulated events from narrow SNR bins, distributed uniformly over a broad range of chirp masses. For nearly identical SNRs, we expect the EoS constraints to be dominated by low-mass events which is due to the following reason. Many candidate EoSs all predict small $\tilde{\Lambda}$ close to 0 for higher mass BNS systems. However, they predict large and often vastly different $\tilde{\Lambda}$ for lighter systems. Thus for larger mass systems, many EoSs in addition to the true EoS are expected to have posterior support since all of them predict $\tilde{\Lambda}$ within a narrow range of 0. On the other hand, for smaller mass systems only the true EoS has support. As an example, see Fig.~\ref{lo-hi}. This leads to larger mass system being much less informative about the EoS than low mass systems with the later dominating the EoS constraints. We verify this by running our algorithm on these sets of of simulated events drawn from narrow SNR bins which yield EoS constraints consistent with this expectation. 

In the next section we describe the single event PE runs which were used to generate the posterior samples that serve as input to our analyses.

\subsection{Single Event EoS-Agnostic Parameter Estimation runs}
\label{PE}
The single event EoS-agnostic PE runs stochastically sample the posterior distribution of parameters $\vec{\theta}$ that characterize the frequency-domain GW waveform model $h(f,\vec{\theta})$, given GW data $p(\vec{\theta}|d)$. By Bayes theorem, this posterior distribution can be expressed as being proportional to the likelihood of obtaining GW data, given said parameters and a noise model, multiplied by uninformative priors on those parameters: $p(\vec{\theta}|d)\propto\mathcal{L}(d|\vec{\theta})~p(\vec{\theta})$. Under the assumption of stationary Gaussian noise~\cite{Bayes0,Bayes1,noise,thranetalbot2019} , the likelihood function is
\begin{eqnarray}
    \mathcal{L}(d|\vec{\theta}) &\propto& \exp[-(d-h(\vec{\theta})|d-h(\vec{\theta}))/2]\label{single-likelihood}\\
    (a|b)&=&4\mathrm{Re}\int_{0}^{\infty}\frac{a^*(f)b(f)}{S_{n}(f)}
\end{eqnarray}
where $S_{n}(f)$ is the noise power spectral density of detector and the frequency domain quantities are obtained by taking a Fourier transform of the time domain quantities in Eq.~\eqref{data0} and  noise. The integral in Eq.~\eqref{single-likelihood} can be evaluated numerically by truncating the waveform model at $f=f_{\text{ISCO}}$, as mentioned in Sec.~\ref{waveform}. For the simulated events, we choose a lower frequncy cutoff at 20\,Hz.  One can now evaluate the posterior $p(\vec{\theta}|d)\propto\mathcal{L}(d|\vec{\theta})p(\vec{\theta})$ and sample it stochastically. For the real events GW170817 and GW190425, we re-use the PE samples released by the LIGO/Virgo Collaboration~\cite{170817PE,190425PE}, that were generated using \textsc{LALInference\_Nest}, of the \textsc{LALSuite} software package~\cite{lalsuite}, which implements nested sampling, to sample the posterior distribution.

However, such nested sampling runs involve potentially millions of likelihood evaluations~\cite{slow4,slow3}, which are computationally costly. Since each likelihood evaluation involves the computation of the entire frequency-domain GW waveform at the corresponding values of $\vec{\theta}$, such a PE run can take potentially weeks per event~\cite{roq1,roq2}. For the purpose of our simulation study, wherein we draw of order 20 events from the Galactic BNS population and of order 10 events each in different SNR and NS mass ranges, we need single event PE analysis techniques that are much more computationally efficient.

For accelerating the analyses, we employ the reduced order quadrature (ROQ) technique \cite{roq1, roq2}. We constructed linear and quadratic ROQ basis vectors of \textsc{TaylorF2} waveform over the parameter space we consider, employing the procedure described in the previous works. To obtain highly compressed basis sets, we constructed tens of linear ROQ basis sets, each of which is constructed over a narrow chirp-mass range, as done in \cite{froq}. The resultant speed-up gain is $\sim10^3$ to $\sim 10^4$, reducing the run time to a few hours.

 We use the \textsc{bilby}~\cite{bilby1,bilby2} package that implements ROQ for single event EoS agnostic PE of simulated BNSs. We choose priors on the BNS parameters for the single event PE runs, that are consistent with previous work. We choose uniform priors in the mass ratio $q\sim U(0,1)$, tidal parameters $\tilde{\Lambda}\sim U(0,5000)$ and $\delta \tilde{\Lambda}\sim U(-5000,5000)$, and the dimensionless component spin parameters $\chi_i\sim U(-0.05,0.05)$. We choose a prior that is uniform in sky position, orientation, and polarization angle. For the real events, the priors used in EoS agnostic PE are  listed in the public release by LVK~\cite{170817PE,190425PE} of the PE samples we use. 
 
 We also use \textsc{bilby}'s implementation of analytic marginalization over extrinsic parameters such as arrival time, coalescence phase and luminosity distance to further accelerate the EoS agnostic PE for the simulated events\cite{thranetalbot2019}. We set uniform priors in time and phase and a powerlaw with index 2 prior on luminosity distance for carrying out the marginalizations. We then exploit \textsc{bilby}'s ability to re-construct posterior samples of the marginalized parameters~\cite{bilby2} in post-processing to obtain posterior samples of luminosity distance which is necessary for EoS inference due to the following reason.
 
 The single event likelihood in Eq.~\eqref{single-likelihood} is implicitly a function of the detector frame chirp mass $\mathcal{M}_d$ which is related to the source frame chirp mass $\mathcal{M}$ by the redshift: $\mathcal{M}_d=\mathcal{M}(1+z)$. Since GW observations alone cannot break the degeneracy between mass and redshift, \textsc{bilby} samples the the likelihood in detector frame chirp mass. However, since the EoS is sensitive to the source frame masses, we need a way to break the mass-redshift degeneracy which is achievable by imposing a cosmological model. Given a cosmological model, posterior samples of luminosity distance (that are reconstructed by \textsc{bilby} in post-processing) can be converted to samples of red-shift. These can then be used to convert posterior samples of detector frame chirp-mass to source frame. Following previous works such as~\cite{gmm}, we use the Planck15 cosmology~\cite{planck15} for converting our chirp masses to detector frame before feeding them into our algorithm for EoS inference. We note that for the low luminosity distances we consider in this study ($<300\,\text{Mpc}$), Planck15 yields redshifts that are small, leading to source frame masses within at most 6-8\% of the detector frame masses.

\vspace{-0.5cm}

\section{Results}
\label{results}
\begin{figure}[!hb]
   \subfigure[GW170817]{\includegraphics[width=0.49\textwidth]{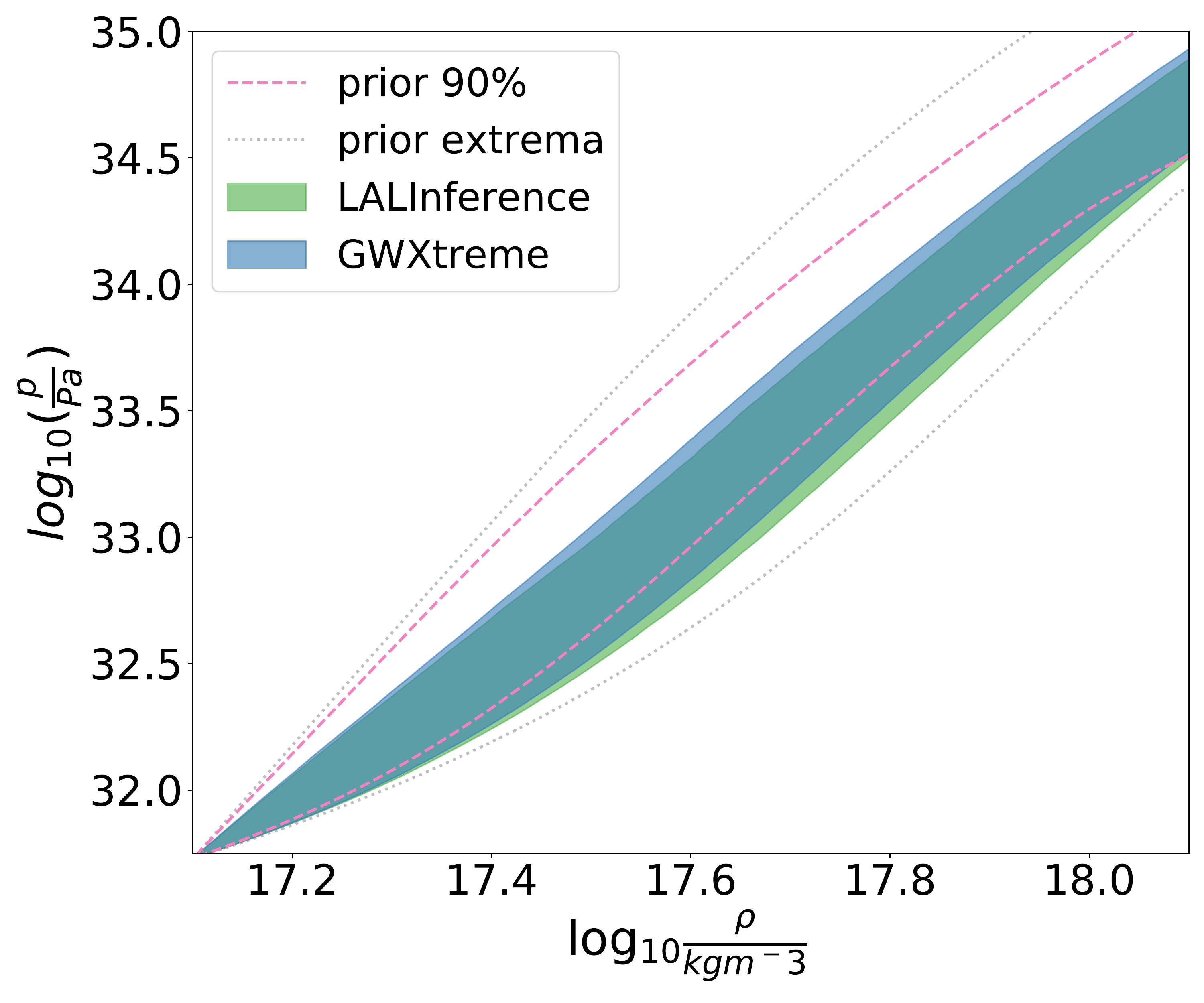}}
    ~
   \subfigure[GW190425]{\includegraphics[width=0.49\textwidth]{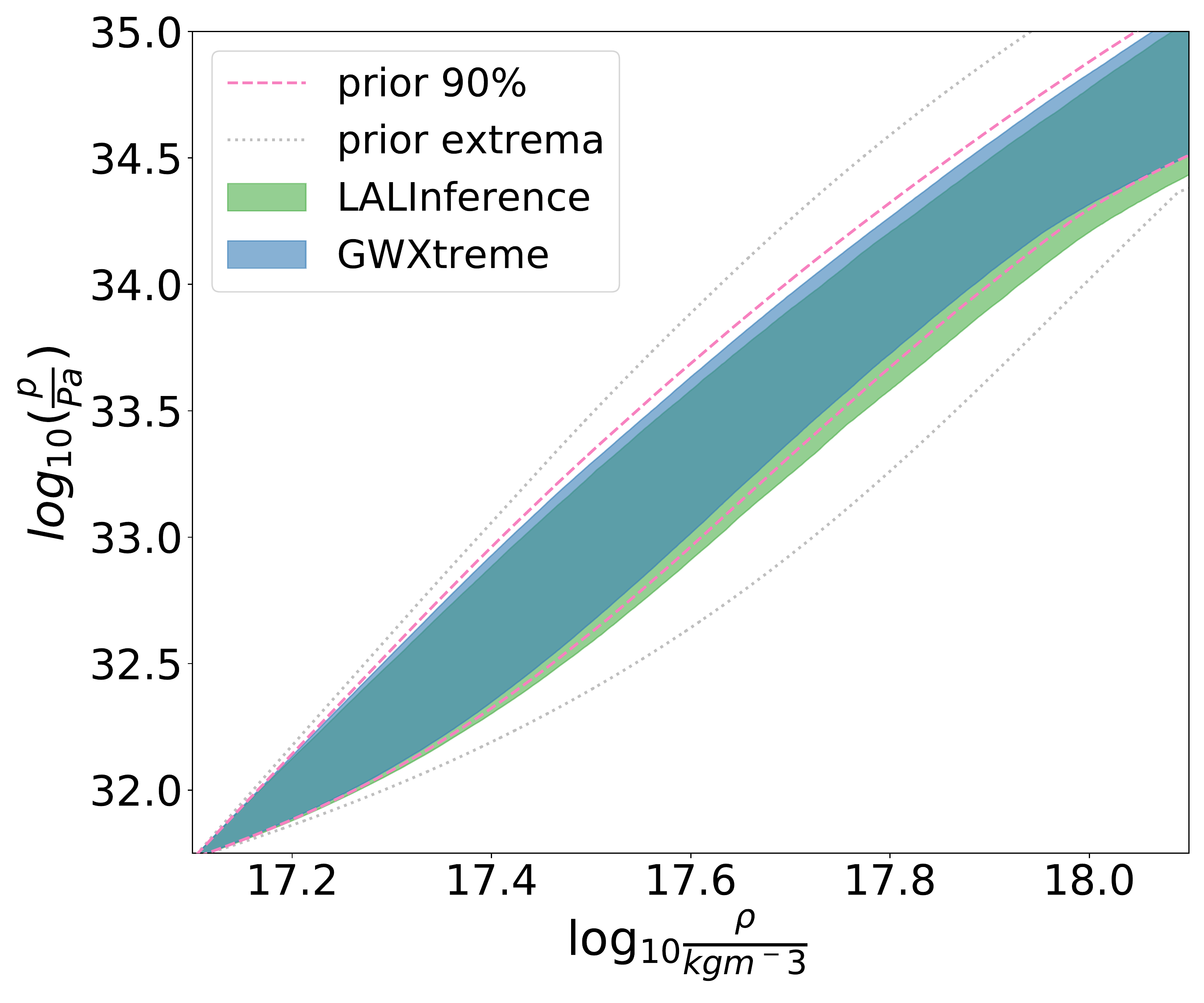}}
    \caption{Comparison of \textsc{GWXtreme} with \textsc{LALInference\_Nest} for the real events GW170817 (left) and GW190425 (right). The posterior samples of the spectral parameters are used to compute pressure density curves and the shaded regions in the plots are equal tail confidence intervals that contain 90 percent of those curves. We also display the 90\% prior intervals in dashed lines and the prior extrema in dotted lines computed using 50000 samples of $\vec{\gamma}$ drawn from the prior. It can be seen that \textsc{GWXtreme} results are consistent with the \textsc{LALInference\_Nest} found by the LVK, results despite being orders of magnitude faster. The slight preference of \textsc{LALInference\_Nest} towards softer EoSs can be attributed to the difference in waveform models as well as the priors on the tidal parameters that are used by the two algorithms.}
    \label{real-data}
\end{figure}
In this section, we display the results we got by running \textsc{GWXtreme}-parameterized on real and synthetic data, which were obtained/generated by using the techniques and details summarized in the previous sections. The algorithm developed based on the method described in Sec.~\ref{approx} is publicly available and documented in the release of \textsc{GWXtreme-0.3.1}. We also reproduce each of these results with the piecewise-polytropic parameterization in the appendix to demonstrate the compatibility of our algorithm with any EoS parametrization.

\subsection{Real Events}

\begin{figure}[h]
    \centering
    \includegraphics[width=0.5\textwidth]{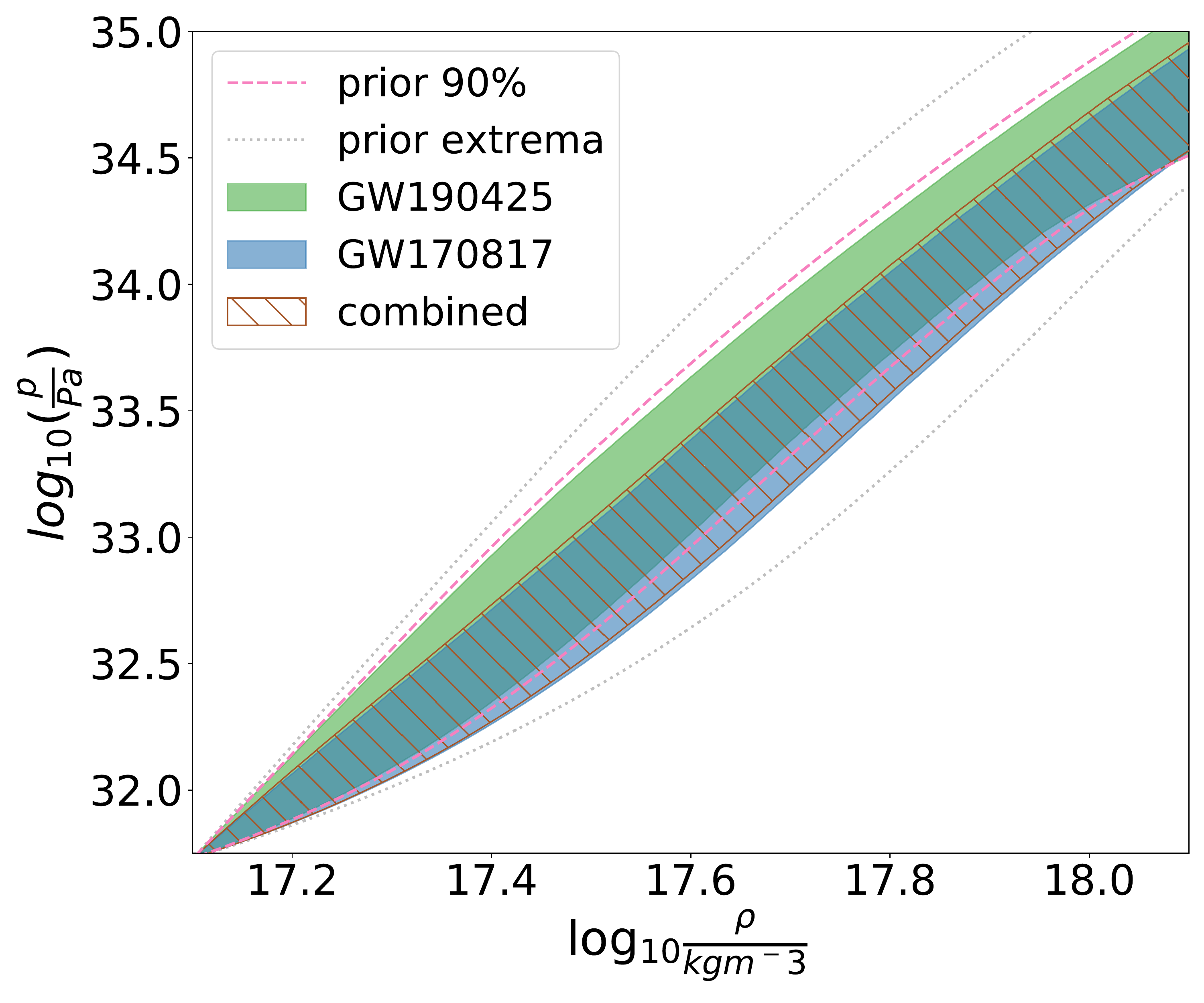}
    \caption{EoS constraints obtained by jointly analyzing GW170817 and GW190425 using  \textsc{GWXtreme}. It can be seen that the joint EoS constraints are dominated by that of GW170817 which is to be expected due its larger SNR and smaller masses of component NSs than GW190425. However, the joint constraint can be seen to be very slightly narrower than that of GW170817 due to the contribution from GW190425.}
    \label{real-data-1719}
\end{figure}
\begin{figure}[!ht]
    \subfigure[EoS Constraint]{\includegraphics[width=0.49\textwidth]{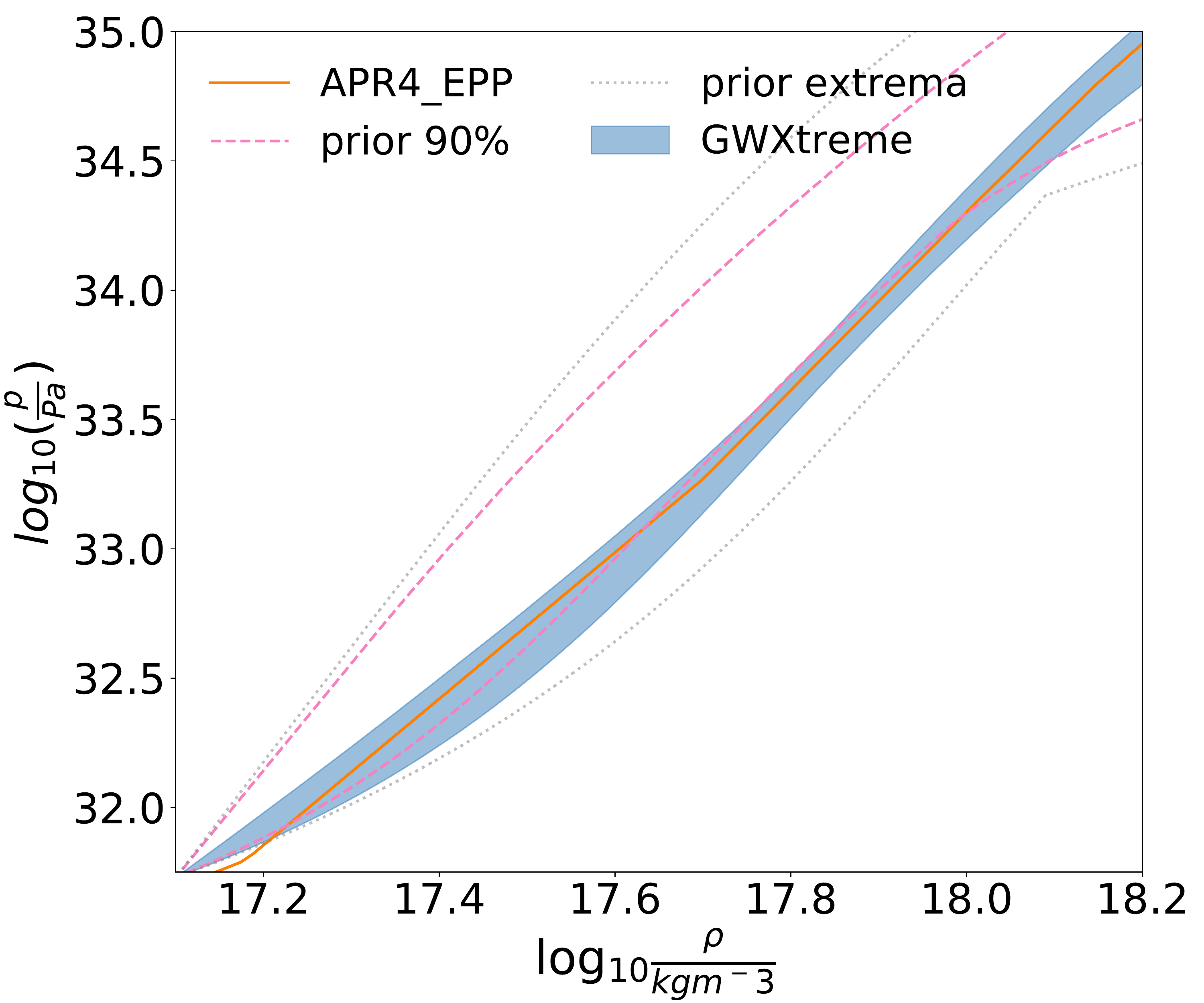}}
    ~
   \subfigure[$\Lambda(1.4M_{\odot})$ constraint]{\includegraphics[width=0.49\textwidth]{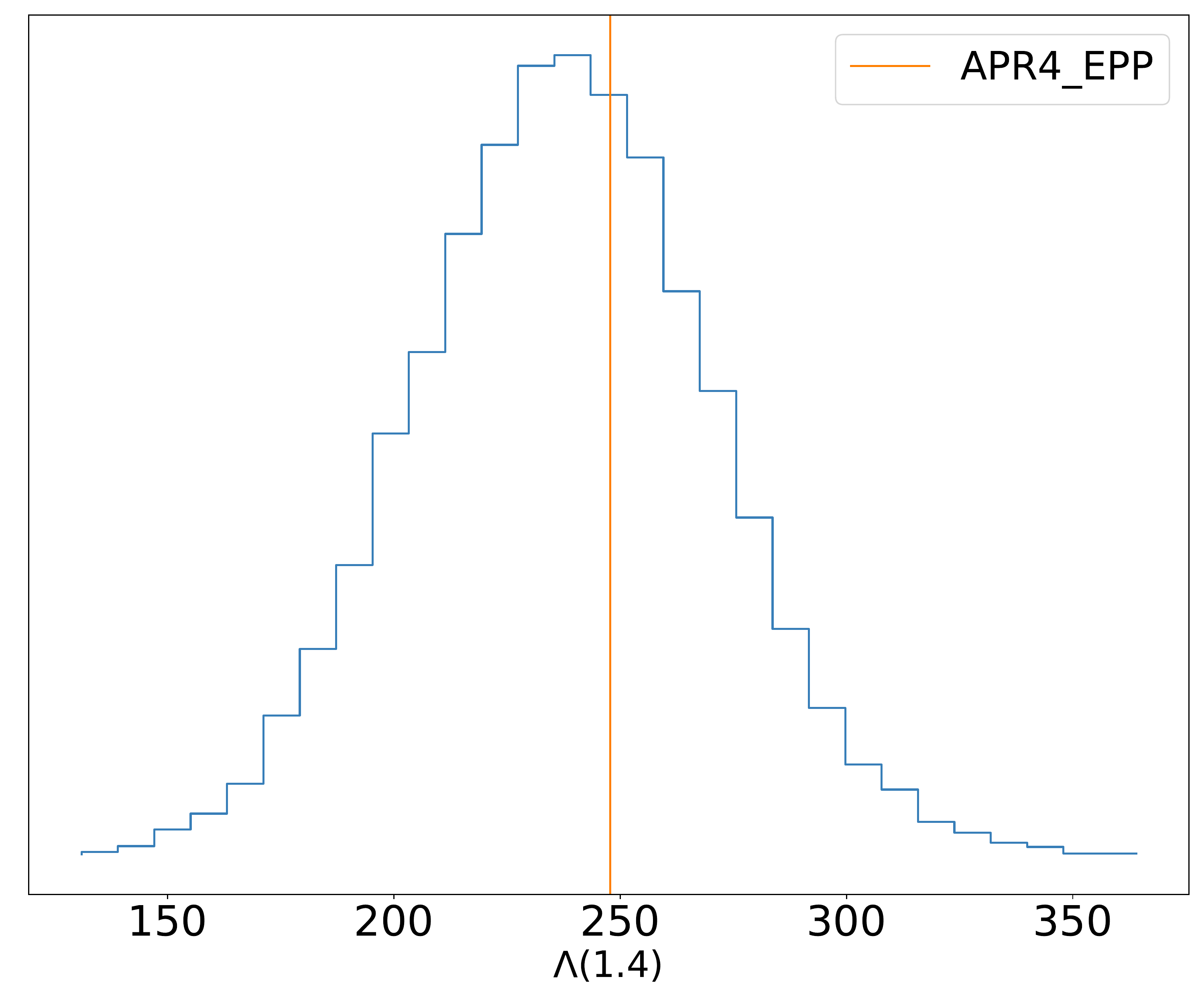}}
    \caption{EoS constraints obtained by jointly analyzing 16 events drawn from the Galactic BNS population with APR4\_EPP as the true EoS and injected into O4 sensitivity. The posterior samples of the spectral parameters generated by \textsc{GWXtreme} can be used to compute $p(\rho,\vec{\gamma})$ and $\Lambda(m,\vec{\gamma})$ curves. In plot (a), the shaded region marks the equal-tail  confidence interval that contains the 90 percent of the $p(\rho,\vec{\gamma})$ curves. Plot (b) is obtained by histogramming $\Lambda(1.4\,M_{\odot},\vec{\gamma}_i)$  corresponding to each posterior sample $\vec{\gamma}_i$. Both plots demonstrate that our computationally cheap and fast algorithm accurately measures the injected EoS.
}
    \label{simulated-data-16}
\end{figure}
\label{real}
We re-used the single event EoS agnostic posterior samples of masses and the tidal parameters given GW data from GW170817 and GW190425, as released by the LVC~\cite{170817PE,190425PE}, that were generated using narrow spin priors and ran our analysis on them to produce EoS constraints in the form of credible intervals on the EoS pressure density plain. We compare our EoS constraints with those obtained by the joint un-approximated PE of EoS hyper-parameters and BNS parameters obtained by the LVK~\cite{170817Eos,190425} using  \textsc{LALInference\_Nest} module of the \textsc{LALSuite} package. Even though different waveform models were used (\textsc{TaylorF2} for the EoS agnostic single event PE for \textsc{GWXtreme}'s input and \textsc{IMRPhenomNRT-v2} for \textsc{LALInference\_Nest}), the EoS constraints are largely consistent, as can be seen  in Fig.~\ref{real-data}. The slightly higher preference of the \textsc{LALInference\_Nest} constraints towards softer EoSs than \textsc{GWXtreme}'s, can be attributed to the difference in the waveform models used in the two analyses. We note that the EoS agnostic posterior distribution of $\tilde{\Lambda}$
 for the events computed using the two waveform models are themselves different. The $\tilde{\Lambda}$ posterior obtained using \textsc{TaylorF2} favors larger $\tilde{\Lambda}$ than \textsc{PhenomNRT} for both GW170817 and GW190425 (see Fig.~11 of~\cite{170817prop} for GW170817 and Fig.~14 of~\cite{190425PE} for GW190415). This explains the preference of stiffer equations of state for \textsc{TaylorF2} and hence \textsc{GWXtreme} which takes the \textsc{TaylorF2}-based EoS agnostic posterior samples of $\tilde{\Lambda}$ as input. From this we can conclude that the slight mismatch in EoS constraints between the two analysis is not due to an artifact of our algorithm but rather due to the difference in waveform models being used. 
 
We also produced a joint constraint by hierarchically combining GW170817 and GW190425 data which we show in Fig.~\ref{real-data-1719}. Due to the large mass of its components and lower SNR as compared to GW170817, the GW190425 event does not contribute much to the joint EoS constraint even though the constraints change slightly from GW170817, which is barely discernible in Fig.~\ref{real-data-1719}.

\subsection{Simulated Events}

\begin{figure*}[!ht]
    \centering
    \subfigure[SNR $\in(23,25)$: EoS Constraint]{\includegraphics[width=0.45\textwidth]{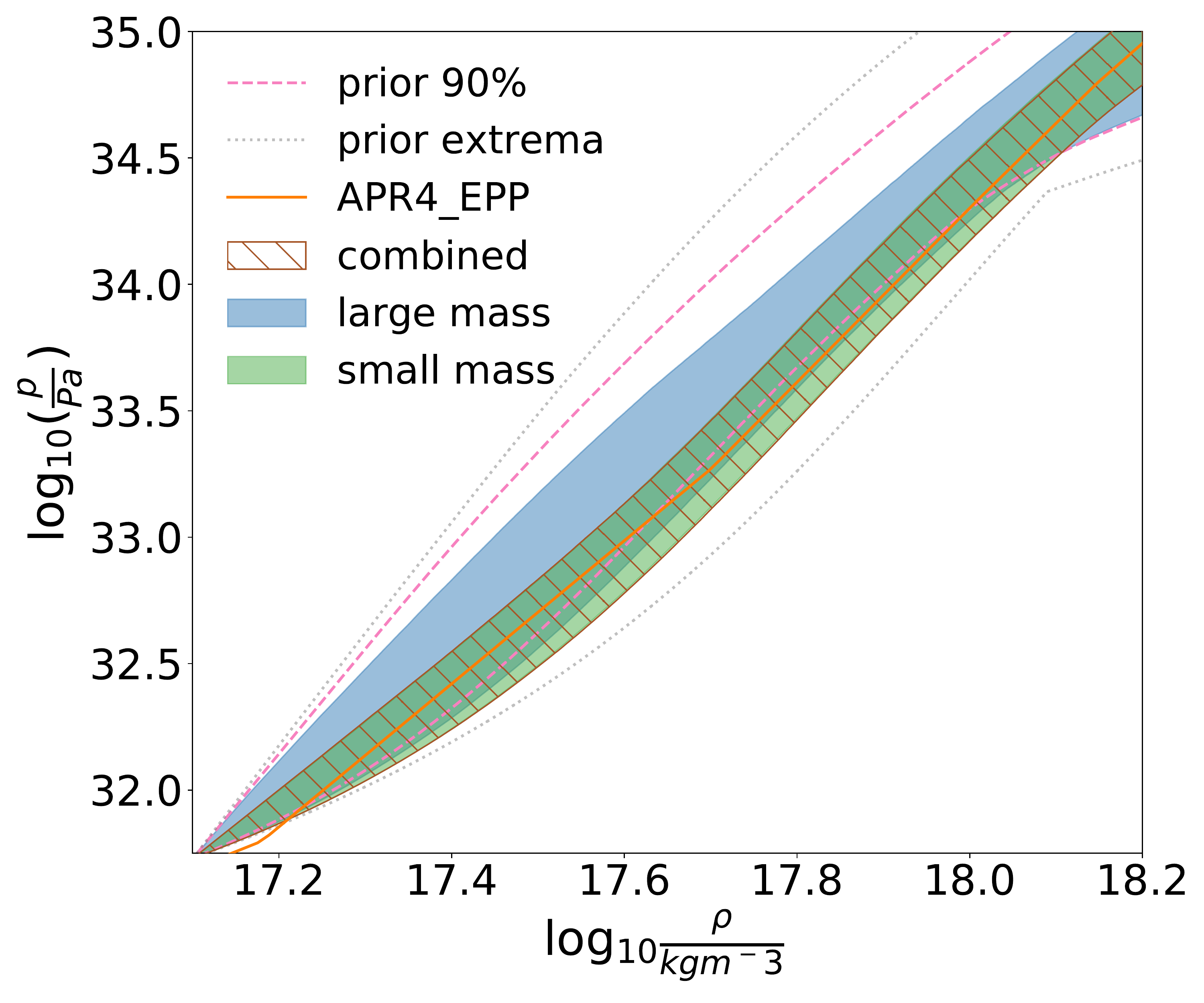}}
    ~
    \subfigure[SNR $\in(23,25)$: $\Lambda(1.4M_{\odot})$ constraint Constraint]{\includegraphics[width=0.45\textwidth]{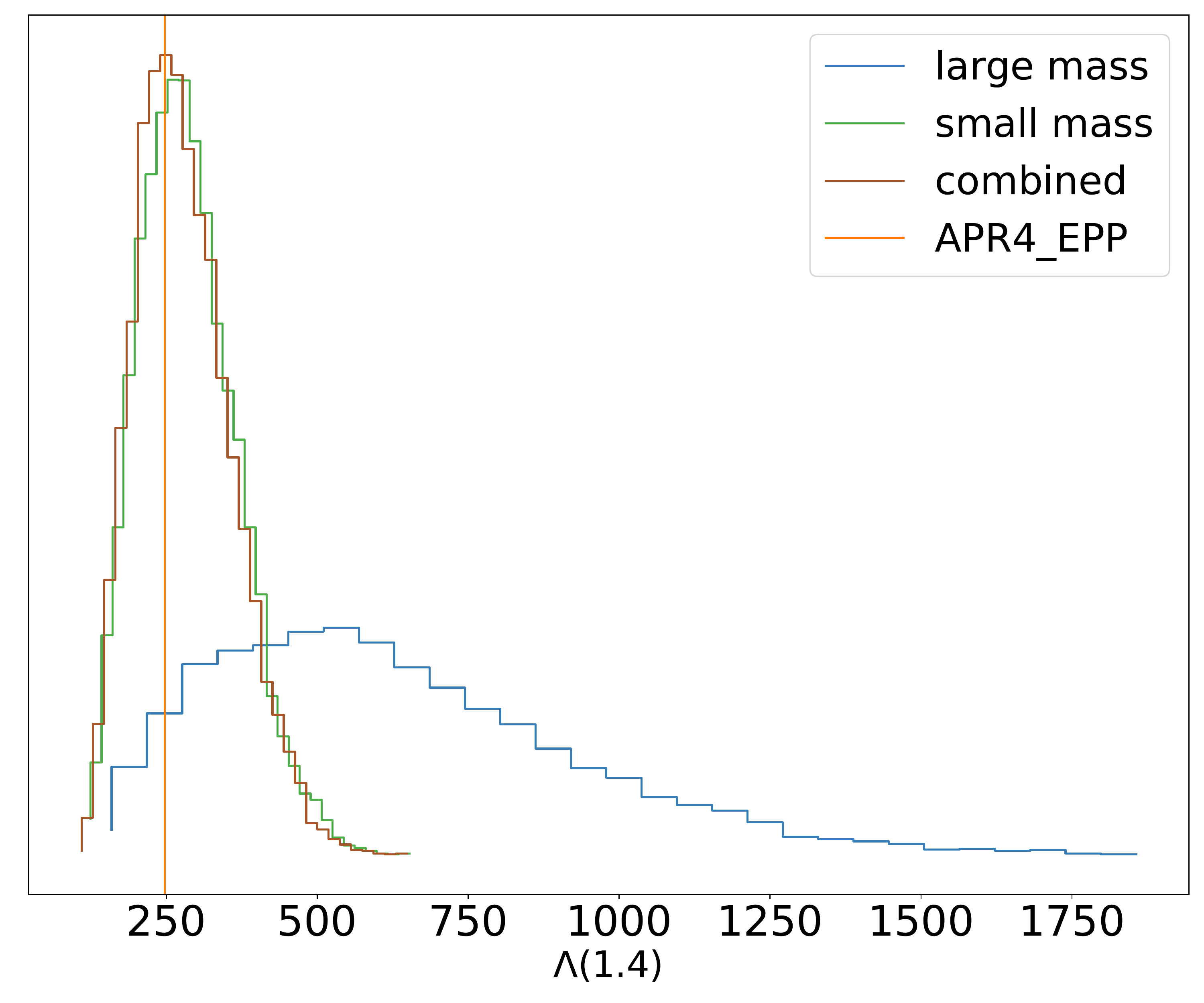}}
    \subfigure[SNR $\in(33,35)$: EoS Constraint]{\includegraphics[width=0.45\textwidth]{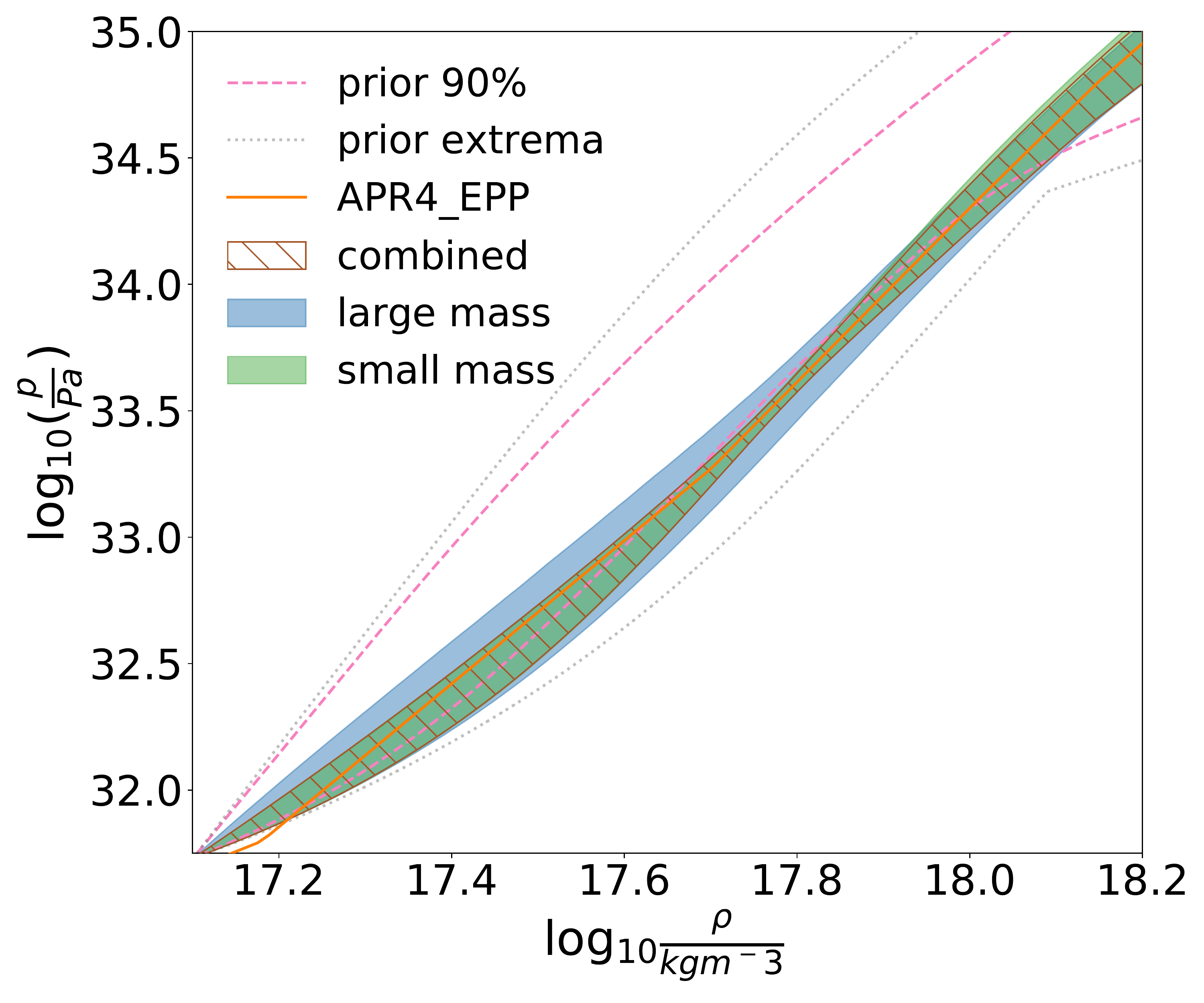}}
    ~
    \subfigure[SNR $\in(33,35)$: $\Lambda(1.4M_{\odot})$ constraint Constraint]{\includegraphics[width=0.45\textwidth]{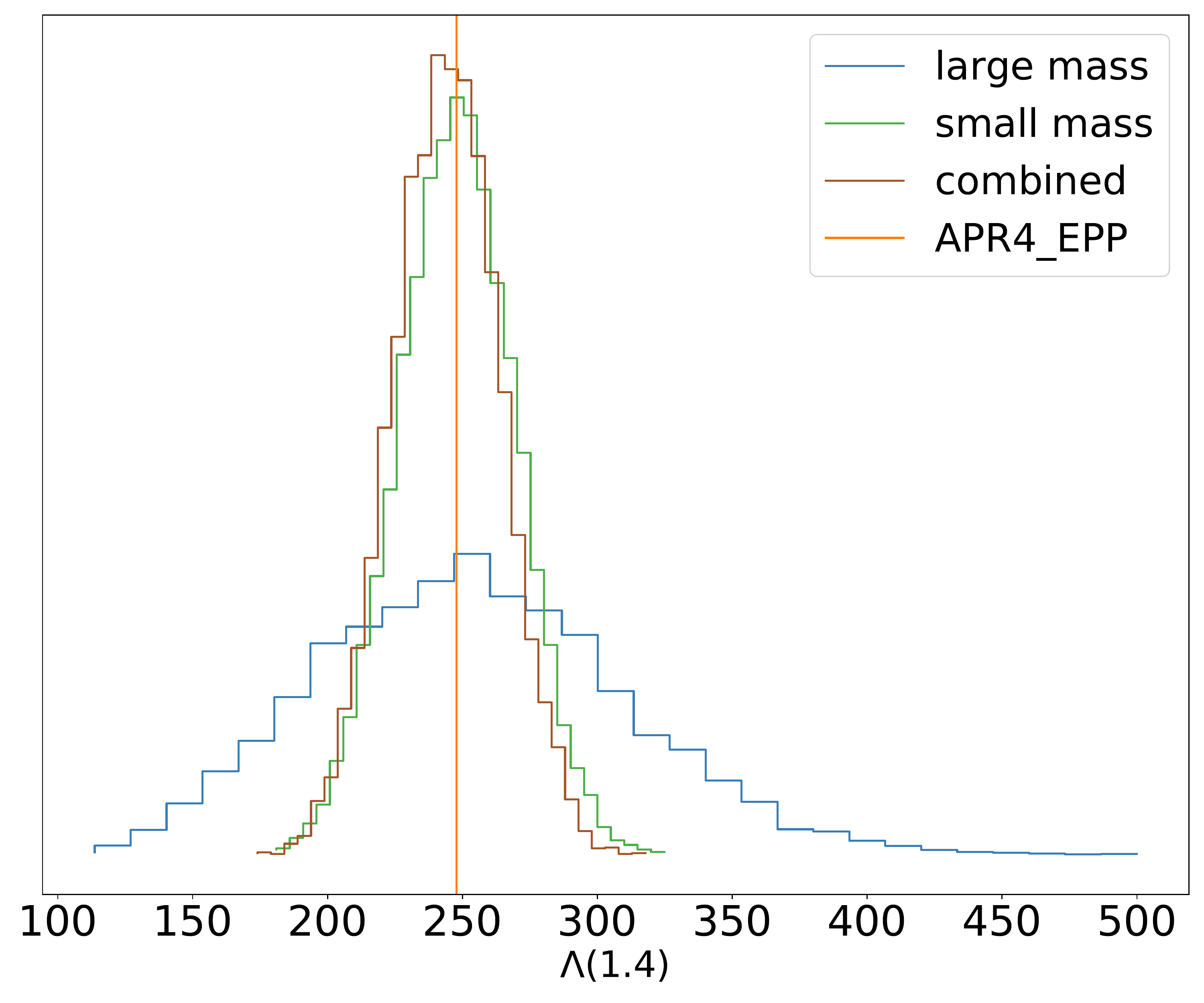}}
    \caption{EoS constraints obtained using \textsc{GWXtreme} for events drawn in narrow SNR ranges. For each chosen SNR range, we draw events uniformly in chirp masses and then draw the extrinsic parameters so that when injected into O4 sensitivity, the simulated signals will have an SNR that belongs in said range of SNRs. We then group the highest mass and lowest mass events in each SNR range and analyze them using our algorithm both separately and jointly. Then using the posterior samples of the spectral parameters generated by our algorithm, we produce EoS constraints as both pressure-density credible intervals and quantile ranges of the tidal deformability at $1.4\,M_{\odot}$. It can be seen that EoS constraints are dominated by the lower mass events in each SNR bin and that they become narrower with increasing SNR.
}
    \label{lo-hi33}
\end{figure*}
\label{sim-art-pop}
In this section we display the results we found upon analyzing the data from 16 simulated events whose masses are distributed according to the Galactic double neutron star population using our algorithm. The details of the simulation and associated modelling of measurement uncertainties are described in Sec.~\ref{data}. Upon analyzing  data generated from such events using our computationally cheap algorithm, we compute EoS constraints that are completely consistent with the chosen ``true'' EoS, as displayed in Fig.~\ref{simulated-data-16}. We note that the increase in the width of the credible intervals near the high and decrease near the low density ends of the EoS pressure density relation are expected given our choice of the parameterized EoS, the population of simulated events and the injected EoS\@. The widening of the constraints at high densities occurs due to those densities being higher than the central density corresponding to the injected EoS, of the most massive component NS in our chosen set of events. On the other hand, the narrowing in the low density end is an artifact of our EoS model. As mentioned in Sec.~\ref{eos-prior} our high density spectral EoS is stitched to SLY at low densities. Due to this stitching with SLY at low densities, variations of the spectral parameters do not affect the pressure density relation at low densities, resulting in the credible interval obtained using the spectral parameters to converge into to a single line (corresponding to SLY) at low densities. We also display the posterior predictive distribution of $\Lambda(1.4\,M_{\odot})$ in Fig.~\ref{simulated-data-16}. The uncertainties in the distribution of $\Lambda(1.4\,M_{\odot})$ can be thought of as estimates of how well \textsc{GWXtreme} can measure the EoS from GW observations and can be used for the comparison of our method with competing ones.

\subsection{Variability of EoS Constraints with SNR and Mass}
\label{sim-bins}
To explore the variability of the EoS constraints with the SNR and chirp mass of events, we simulated events randomly in narrow SNR bins of $(23,25)$ and $(33,35)$, with the chirp masses for each bin chosen uniformly in the range $(0.8\,M_{\odot},1.8\,M_{\odot})$.
The joint EoS constraints along with the predictive distributions of the measured dimensionless tidal parameter $\Lambda(1.4\,M_{\odot})$, obtained from analyzing these events using our algorithm is displayed in Fig.~\ref{lo-hi33}. For the $(23,25)$ SNR range we see that the high mass events are much less informative than the low mass once and that the joint constraint is completely dominated by the lower mass events even in the high density regime. This is consistent with what we expect, as described in Sec.~\ref{data}. We see similar trends in the $(33,35)$ SNR range along with a couple of additional features. First, both high mass  and low mass events produce narrower constraints than their $(23,25)$ SNR counterparts. Second, in the high density regime, the joint constraint appears more informative than the one obtained from low mass events only due to contribution from the high mass events. This implies that in joint EoS inference from multiple events, a high mass event can result in a non-neglible information gain only if it is loud enough.

\section{Conclusion and Future Prospects}
\label{conclusion}
We have developed an algorithm for fast and computationally cheap hierarchical inference of the NS EoS using observations of GWs from multiple BNSs that re-uses single event EoS agnostic PE results to achieve its latency and efficiency. We demonstrated the accuracy of our method by showing its results to be fully consistent with the existing EoS constraints for the events GW190425 and GW170817 that were computed using un-approximated and hence much costlier analyses. We also demonstrated the accuracy with which our method can constrain the true EoS by performing a simulation study with realistic modeling of the measurement uncertainties in the EoS sensitive BNS parameters and a population of BNSs consistent with the latest studies of the Galactic double neutron star populations. We further studied the variability of EoS constraints with BNS chirp mass and SNRs by drawing simulated events in narrow SNR bins and over a broad range of BNS chirp masses. We found that EoS constraints are dominated by lower mass and higher SNR events, in agreement with our understanding of NS structure, which can be used for potentially truncating the list of events that need to be analyzed for joint EoS constraints without losing precision. While the variation with SNR is consistent with previous works, we have shown that the variation with chirp mass is also equally significant in selecting the events that will have the most dominant contribution to the EoS constraints.

Even though we do not simultaneously infer the BNS mass distributions which might have non-negligible corelations with the EoS constraints~\cite{interpolation}, we note that our method is generalizable to do so. Simultaneous inference of BNS mass distributions in our framework would require the uninformative prior $p(\mathcal{M},q)$ in Eq.~\eqref{likelihood3} to be replaced by a population prior conditional on the mass population model and its parameters that we are trying to infer: $p(\mathcal{M},q|\vec{\gamma}_{\text{pop}})$, where $\vec{\gamma}_{\text{pop}}$ are the universal parameters characterizing the BNS mass distribution. This will induce an additional factor of $p(q,\bar{\mathcal{M}}|\vec{\gamma}_{\text{pop}})/p(\vec{q},\bar{\mathcal{M}})$ in the integrand of Eq.~\eqref{likelihood31} and hence those of Eq.~\eqref{posterior2}. Then, Including an additional prior on the mass population parameters $p(\vec{\gamma}_{\text{pop}})$ to be multiplied with the prior on the EoS parameters $p(\vec{\gamma}|I)$ in Eq.~\eqref{posterior2} effectively makes its LHS the joint posterior of the universal EoS and population parameters given GW data from multiple events, which can be sampled stochastically to produce simultaneous mass-population and EoS constraints. Since the KDE and rest of the approximations along with the dimensionality of the integrals remain the same, this generalization will not affect the latency and computational cost of our algorithm. We leave such a generalization as an upcoming work.

We have shown that our algorithm is compatible with the \textsc{TaylorF2} waveform model. We note that switching to a 3-dimensional KDE in $(\Lambda_1,\Lambda_2,q)$ instead of our 2 dimensional one, has the potential of making our analysis compatible with other waveforms as well, such as \textsc{PhenomPNRT}. As described in Secs.~\ref{waveform} and~\ref{approx}, our 2-dimensional KDE-based approximation necessitates the use of a uniform in $\tilde{\Lambda}$ prior in the single event EoS agnostic PE runs which prevents the use of existing PE results with the \textsc{PhenomPNRT} waveforms. However the 3-dimensional KDE based generalization would work with the uniform in $\Lambda_1$-$\Lambda_2$ prior being used in the single event PE with \textsc{PhenomPNRT} waveforms, enabling the use of these PE results. For such a prior, i.e., $p_{\text{PE}}(\Lambda_1,\Lambda_2,\mathcal{M},q)\propto p(\mathcal{M},q)$, the 3-dimensional KDE approximation to the EoS agnostic posterior, $p(q,\Lambda_1,\Lambda_2|d_i)\approx K_i(q,\Lambda_1,\Lambda_2)$, would lead to the modification of Eq.~\eqref{posterior2} to $p(\vec{\gamma}|\{d\},\mathcal{E},I) \appropto p(\vec{\gamma},I)\prod_{i=1}^N \int_{0}^1 K_i(q,\Lambda_{1}(\bar{\mathcal{M}}_i,q,\vec{\gamma}),\Lambda_{2}(\bar{\mathcal{M}}_i,q,\vec{\gamma}))\,dq$. This posterior can be then be sampled stochastically using the same techniques outlined in Sec.~\ref{approx} to produce EoS constraints. Similar calculations with higher dimensional KDE's have been shown to produce sensible results in non-parametric EoS inference studies like refs~\cite{3dkde1,3dkde2,3dkde3}. In the context of our algorithm, such a calculation is a generalization we leave as part of future work. 

To summarize, in this proof of concept work, we develop an algorithm for fast and efficient hierarchical Inference of parameterized NS EoS from multiple GW observations and demonstrate its accuracy. With this development, \texttt{GWXtreme} is now a strong candidate for performing fast and computationally cheap hierarchical EoS inference using both tabulated and parameterized EoS models from multiple GW obsevations in O4. We have noted that generalizations of our algorithm to increase accuracy and applicability while maintaining efficiency is straightforward and will be available soon in future releases of \textsc{GWXtreme}. We further note that our demonstrations serve as proof of concept for the applicability of bounded KDEs in increasing the efficiency of other GW based hierarchical inference problems. Similar problems where-in the parameterized physical model being inferred implies deterministic relationships between event-specific observables leading to delta function priors on them, can be efficiently handled with customized KDEs. Our algorithm can serve to guide such analyses which can re-use the basic concept of our framework while needing to modify only the model-imposed priors and set of observables sensitive to the model.

\acknowledgments
The authors would like to thank Leslie Wade and Ignacio Maga\~{n}a Hernandez for useful discussions. The authors would like to thank Philippe Landry for valuable suggestions and for conducting the internal review. This work was supported by the National Science Foundation awards PHY-2207728 and PHY-2110576. The authors are grateful for computational resources provided by the LIGO Laboratory and supported by National Science Foundation Grants PHY-0757058 and PHY-0823459, PHY-0823459, PHY-1626190 and PHY-1700765. This material is based upon work supported by NSF's LIGO Laboratory which is a major facility fully funded by the National Science Foundation

\appendix

\section{Validation Study: The Piecewise-Polytropic Parametrization}
\label{appendix-piecewise}
In this appendix, we demonstrate that our algorithm is compatible with potentially any EoS parameterization by reproducing our results with the 4-parameter piecewise-polytropic parameterization instead of the spectral one. We run our algorithm with the piecewise parameterization on the same data for which we displayed the spectral results in Sec.~\ref{results}. The piecewise polytropic parameterization is based on the assumption that the EoS pressure density relation can be accurately represented by multiple polytropes (power-law relations of the form $p=K\rho^{\Gamma}$), stitched together at fixed joining densities. If the joining densities $\rho_i$ are fixed then such a parameterization containing $n_p$ polytropes can be completely characterized by $n_p+1$ parameters: the adiabatic indices $\Gamma_i$ of each polytrope and the pressure at the first joining density $p_1$. 

For a four parameter model, three polytropes are stitched together resulting in an EoS characterized by the free parameters: $\vec{\gamma}=\{\log{p_1},\Gamma_1,\Gamma_2,\Gamma_3\}$. We impose priors on these parameters that are physically equivalent to the priors imposed on the spectral parameters as described in Sec.~\ref{eos-prior}. We impose the thermal stability condition by imposing a uniform prior on the parameters: $\log_{10}(p / \text{dyn}\,\text{cm}^{-2}) \in [33.6,35.4]$, $\Gamma_1\in[2,4.5]$,$\Gamma_2\in[2.0,4.5]$ and $\Gamma_2,\Gamma_3\in[1.1,4.5]$, which are also broad enough to accurately fit a large number of candidate EoS models. In addition to these uniform bounds, we further impose the causality of sound speed and the observational consistency of maximum NS mass prior exactly similar to the spectral case, as described in \ref{eos-prior}. These choice of priors are consistent with previous works such as~\cite{piecewiseonly} and~\cite{spectralpiecewise}. 

\begin{figure*}[!ht]
    \centering
    \begin{subfigure}
        \centering
        \includegraphics[width=0.45\textwidth]{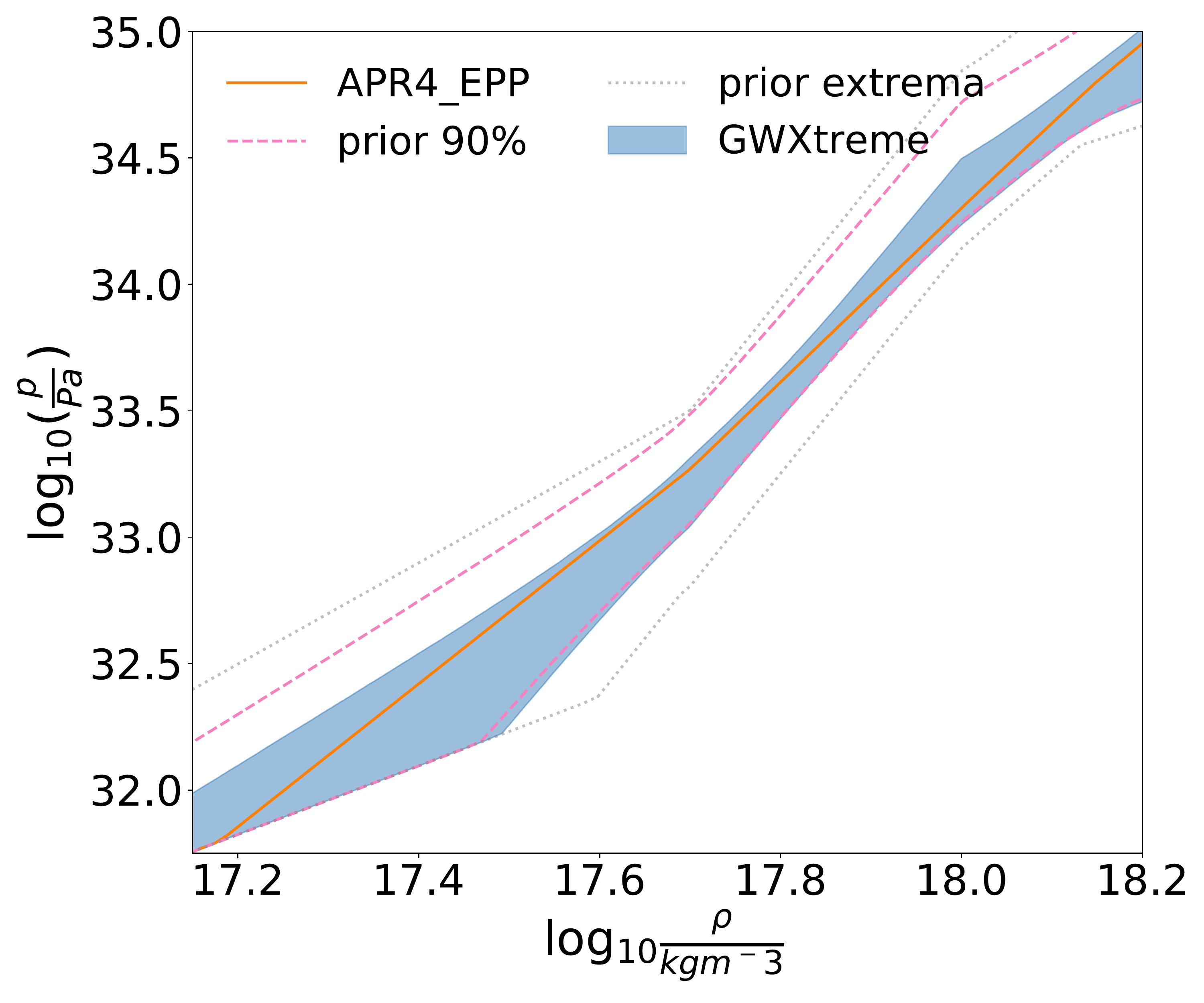}{a}
    \end{subfigure}%
    ~
    \begin{subfigure}
        \centering
        \includegraphics[width=0.45\textwidth]{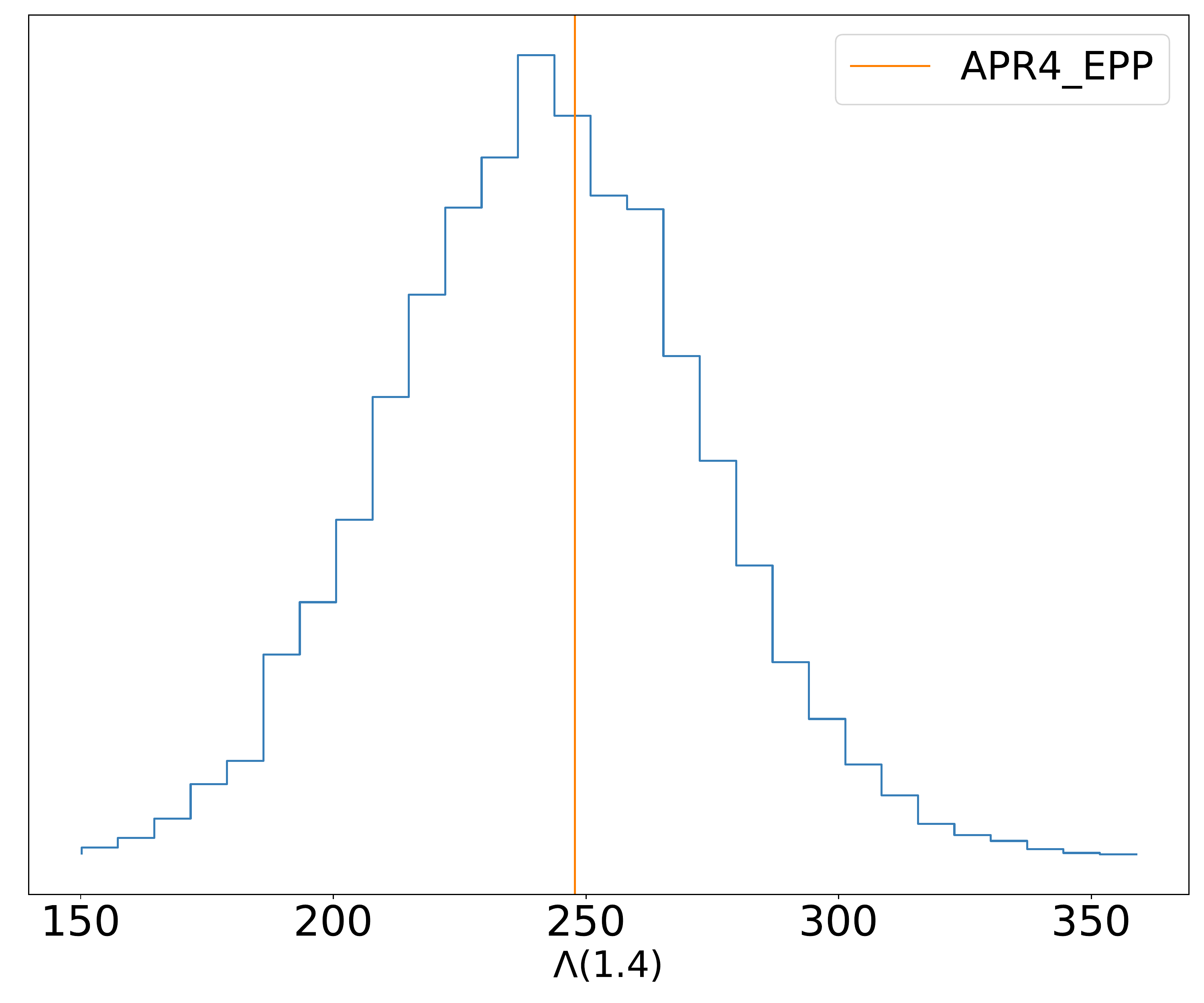}{b}
        
    \end{subfigure}
    \caption{EoS constraints obtained using the piecewise polytrope  for the 16 events drawn from the galactic population distribution
}
    \label{piecewise-16}
\end{figure*}

\begin{figure*}[!ht]
    \centering
    \begin{subfigure}
        \centering
        \includegraphics[width=0.45\textwidth]{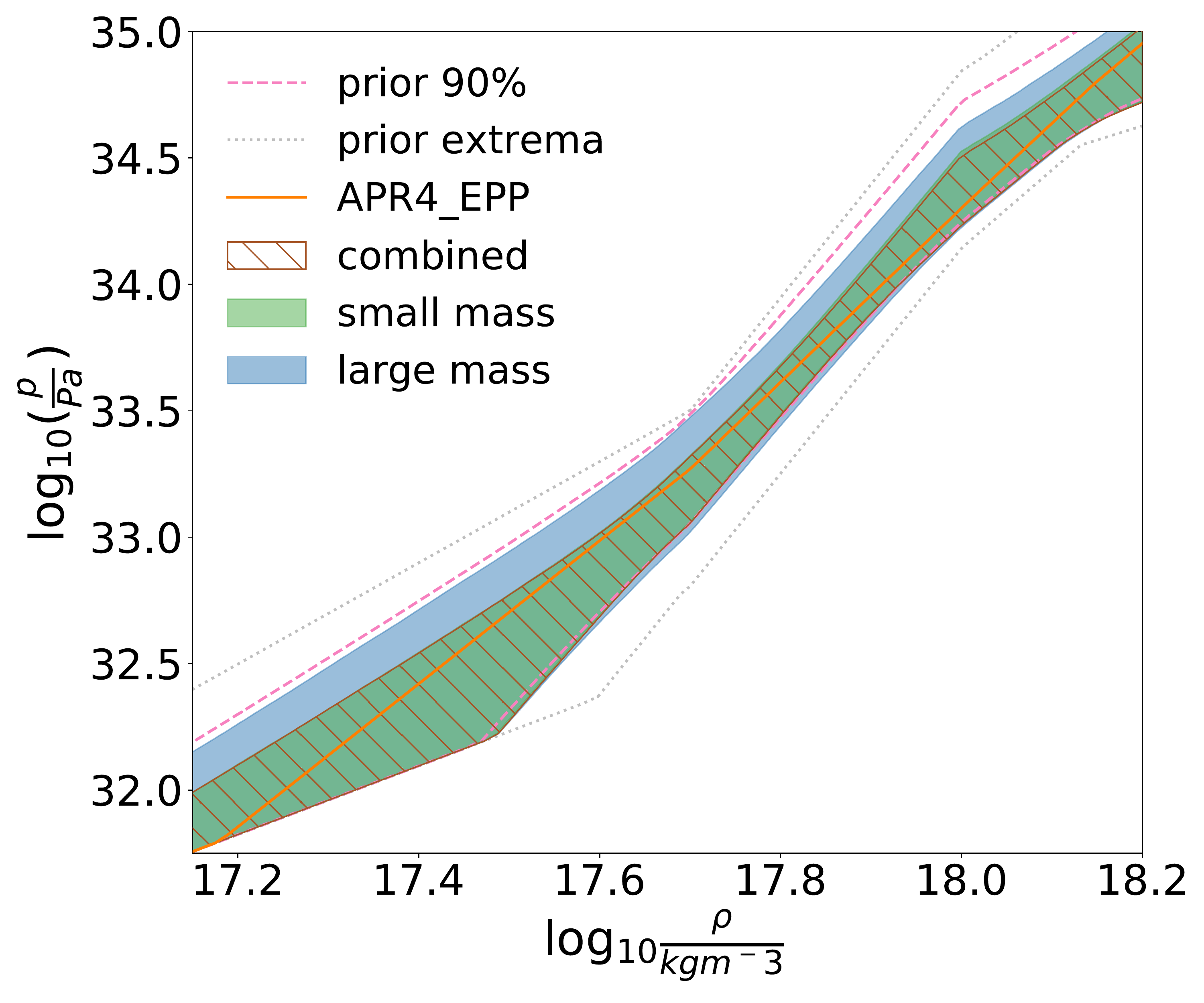}{a}
    \end{subfigure}%
    ~
    \begin{subfigure}
        \centering
        \includegraphics[width=0.45\textwidth]{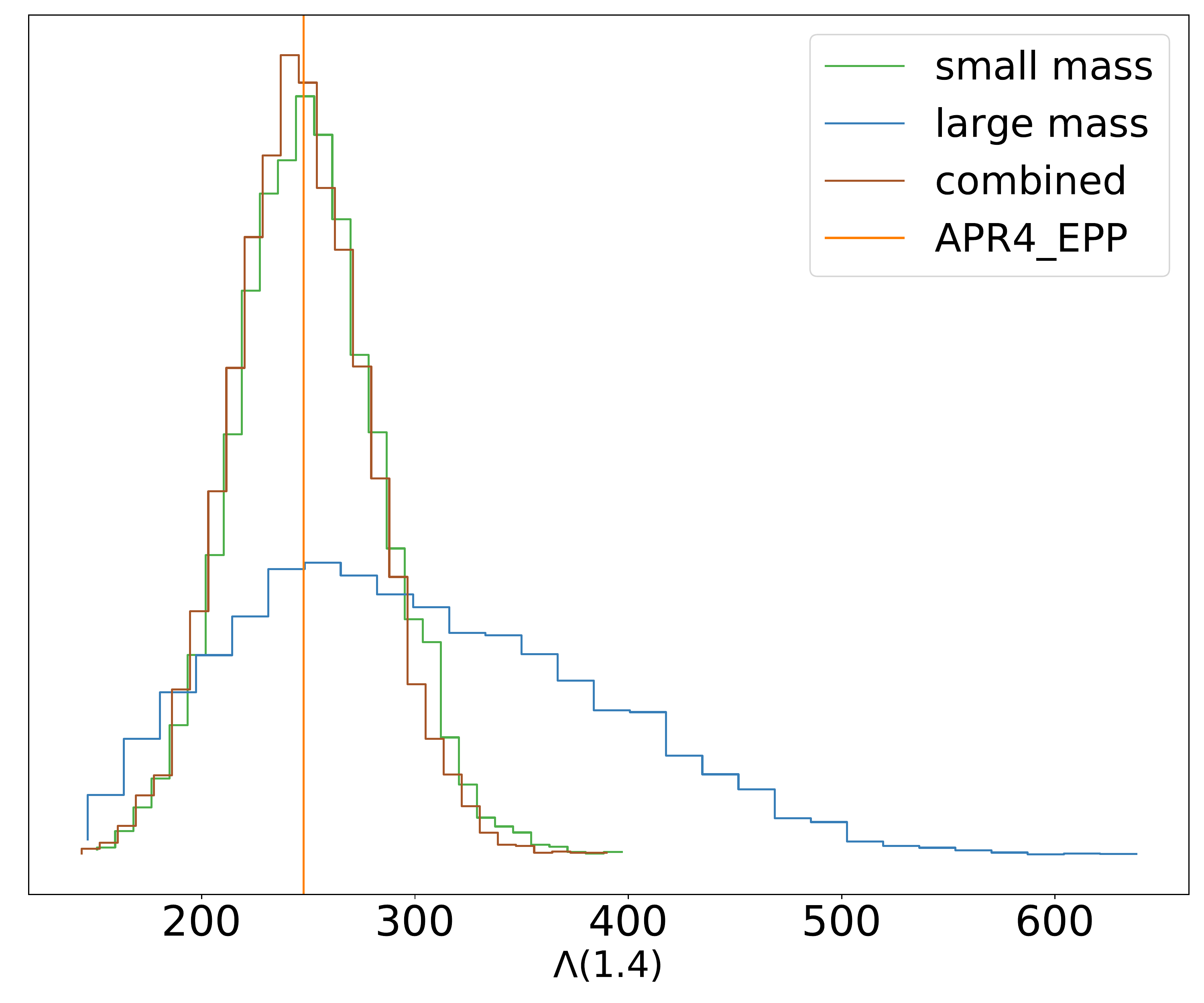}{b}
    \end{subfigure}
    \caption{EoS constraints obtained using the piecewise polytrope  for simulated events drawn in the 23 to 25 SNR bin
}
    \label{piecewise-23-25}
\end{figure*}

 We again use \textsc{LALSimulation} to solve the TOV equations for this four parameter realization of the piecewise-polytropic parameterization to convert pressure-density relation into a mass and tidal parameter relation. As implemented in \textsc{LALSimulation}, we choose the joining densities $\rho_1$ and $\rho_2$ to be $\rho_1=1.8\rho_{\text{nuc}}$ and $\rho_2=3.6\rho_{\text{nuc}}$ respectively, where $\rho_{\text{nuc}}\approx 2.8\times10^{15}\,\text{g}\,\text{cm}^{-3}$ is the nuclear saturation density. As in the case of the spectral parameterization, we anchor the high density parameterized EoS with a fixed low density EoS, which is chosen to be the SLY. This implementation of the piecewise polytropic parameterization is consistent with previous works. Under these considerations, with the priors described above, running our algorithm for the simulated events yields EoS constraints that are completely consistent with the spectral results as can be seen in Figs.~\ref{piecewise-16} to~\ref{piecewise-23-25}. 

\begin{figure*}[!ht]
    \centering
    \begin{subfigure}
        \centering
        \includegraphics[width=0.45\textwidth]{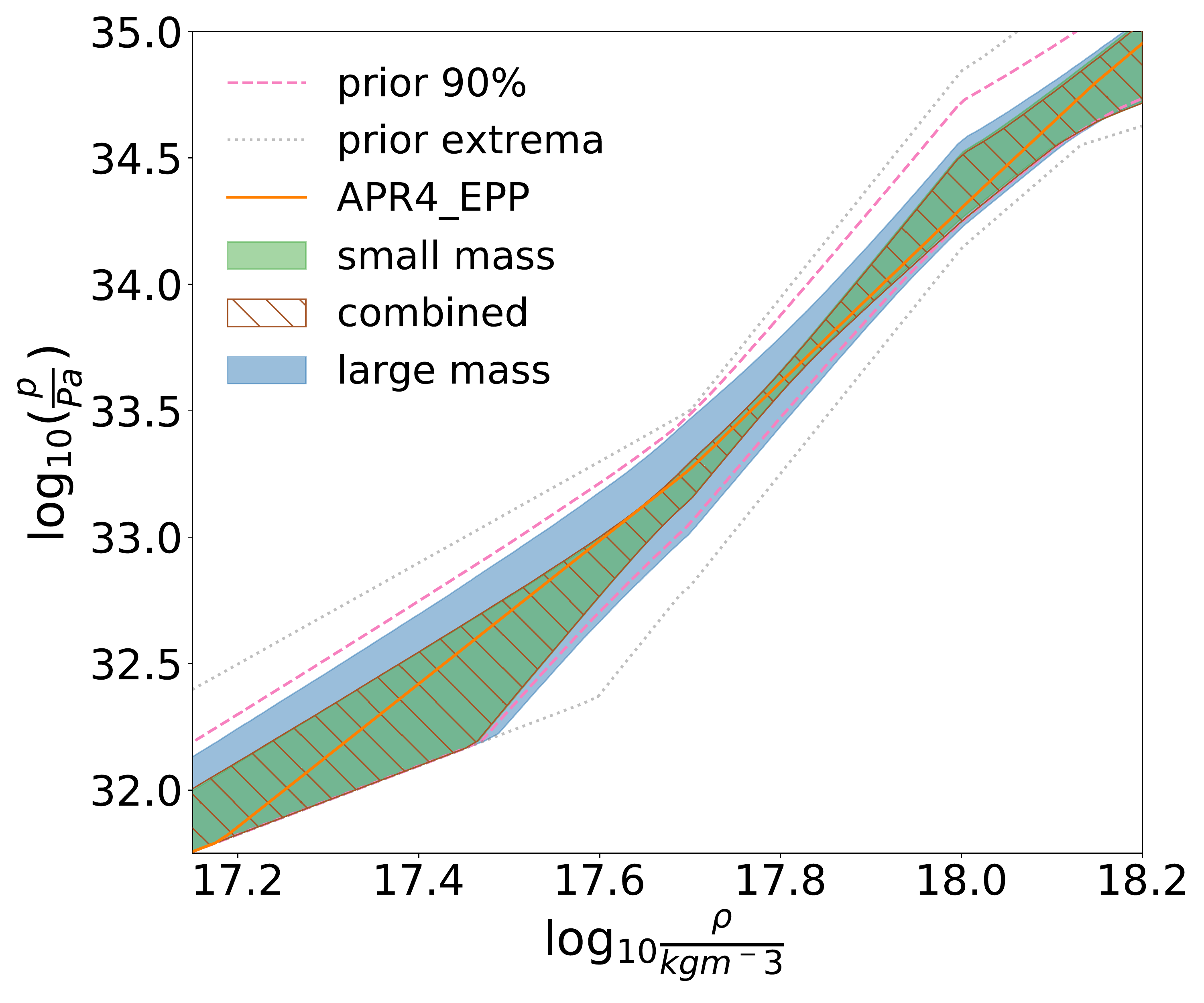}{a}
    \end{subfigure}%
    ~
    \begin{subfigure}
        \centering
        \includegraphics[width=0.45\textwidth]{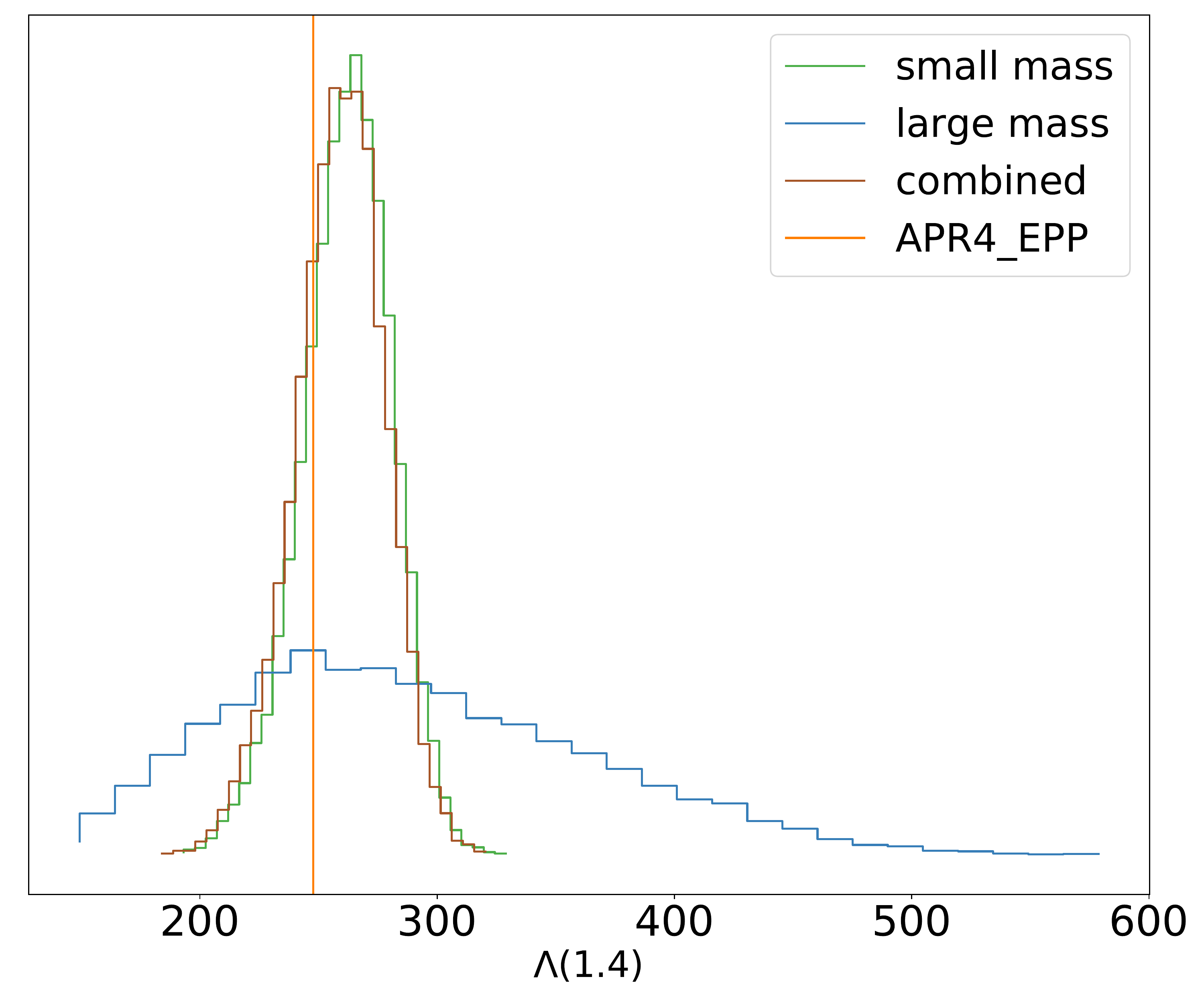}{b}
    \end{subfigure}
    \caption{EoS constraints obtained using the piecewise polytrope  for events drawn in the 33 to 35 SNR bin
}
    \label{piecewise-33-35}
\end{figure*}
\vspace{-0.5cm}
\subsection{Comparison with the spectral parameterization}
While these results in Fig.~\ref{piecewise-16} through Fig.~\ref{piecewise-33-35} are in complete agreement with the injected EoS and what we expect in terms of the variability of the EoS bounds with mass and SNR, we can see sharp changes in the EoS constraints near the joining densities of the polytrope pieces. Near those densities, we can also identify broadening of the EoS constraints which can be attributed to the non-differentiability of the EoS at those points and hence interpreted as artificial. This pathology is inherent to the piecewise parameterization, and is absent in the spectral parameterization.

Due to these aforementioned issues with the piecewise parameterizations we chose the spectral EoS for our main result, leaving the results of the piecewise polytropic parameterization as a validation study, that demonstrates the compatibility of our algorithm with multiple EoS parameterizations. The exact origin of these issues with the piecewise parameterization is an interesting academic exercise that we do not pursue in greater detail since the spectral EoS, which is free of these issues, has been shown to produce consistent results in conjunction with our algorithm.

\section{Number of Simulated Events}
\label{poisson}
The number of events one might expect to observe in O4 can be estimated from the number of events seen up until now and the ratio of spacetime hypervolume that will be surveyed in O4 to the hypervolume previously surveyed.  Due to the small number of BNSs observed to date, this estimated number will have a large uncertainty. Here a simple calculation using Poisson statistics is used to make reasonable guesses for the expected number of events, within counting uncertainties. Let the observed rate of BNS events during a past observing be $\mu_0$. Then the likelihood of observing $n_0$ number of events during a past observing run given $\mu_0$ can be modeled by a Poisson distribution:
\begin{equation}
    p(n_0|\mu_0) = \frac{1}{n_0!}\mu_0^{n_0} e^{-\mu_0}\label{rate1}
\end{equation}
Using Bayes theorem and some uninformative prior on $\mu_0$ such as a powerlaw, $p(\mu_0)=\mu_0^{-\beta}$, one can write the posterior distribution of the true rate given the number of past observations $n_0$, after the correct normalization becomes
\begin{equation}
    p(\mu_0|n_0) = \frac{1}{\Gamma(n_0+1-\beta)}\mu_0^{n_0-\beta}e^{-\mu_0}\label{rate2}
\end{equation}
If the new observing run is $\alpha$ times more sensitive than the past one then the rate of observed events in the new run relates to that of the older run via $\mu=\alpha \mu_0$. Substituting into Eq.~\eqref{rate2} allows us to compute the probability of the observed rate during the new run given the number of observations in the older run: $p(\mu|n_0)$. Noting that the expected number of events $n$ to be observed during the new observing run given $\mu$ is another Poisson distribution like Eq.~\eqref{rate1}: $p(n|\mu)=\mu^ne^{-\mu}/n!$, one can marginalize over $\mu$ and find the probability of observing $n$ given the observed $n_0$:
\begin{widetext}
\begin{equation}
    p(n|n_0) =  \int_0^{\infty} p(n|\mu)p(\mu|n_0)d\mu=\frac{\Gamma(n+n_0+1-\beta)}{\Gamma(n+1)\Gamma(n_0+1-\beta)} \frac{\alpha^n}{(1+\alpha)^{n+n_0+1-\beta}}\label{p(n)}
\end{equation}
\end{widetext}
A sanity check of this equation is to show that the total probability, when summed over $n$, is unity, or, equivalently, that
\begin{equation}
    (1+\alpha)^{n_0+1-\beta} = \sum_{n=0}^\infty \frac{1}{n!} \frac{\Gamma(n+n_0+1-\beta)}{\Gamma(n_0+1-\beta)}\left(\frac{\alpha}{1+\alpha}\right)^n.
\end{equation}
To show this, consider the function $(1-z)^{-a}$ for real $a>0$, which is analytic in the domain $|z|<1$.  Its Maclauren series is
\begin{equation}
    \frac{1}{(1-z)^a}=\sum_{n=0}^\infty \frac{1}{n!}\frac{\Gamma(a+n)}{\Gamma(a)}z^n.
\end{equation}
Now let $z=\alpha/(1+\alpha)$ (note: $|z|<1$ for $\alpha>0$) and $a=n_0+1-\beta$ (note: $a>0$ for $n_0\ge0$ and $\beta<1$), which completes the proof.
As a second sanity check, the expected number of events, $\langle n\rangle$, is computed by differentiating both sides of
\begin{equation}
    1 = \sum_{n=0}^\infty \frac{\Gamma(n+n_0+1-\beta)}{\Gamma(n+1)\Gamma(n_0+1-\beta)} \frac{\alpha^n}{(1+\alpha)^{n+n_0+1-\beta}}
\end{equation}
with respect to $\alpha$ and then multiplying by $\alpha(1+\alpha)$; the result is
\begin{equation}
    \langle n \rangle = \sum_{n=0}^\infty n p(n|n_0) = (n_0+1-\beta)\alpha
\end{equation}
which, as expected, is proportional to $\alpha$.  Repeating the procedure on the above equation results in $\mathord{\mathrm{Var}}\,n=\langle{n^2}\rangle-\langle n\rangle^2=(\alpha+1)\langle n\rangle$.

We note that the ratio $\alpha$ that relates the observed rates of two different runs is essentially the ratio between the spacetime hypervolume $\langle VT\rangle$ to which these runs are sensitive. Since no data is available for computing $\langle VT\rangle$ of O4, we approximate $\alpha$ from the effective BNS ranges of the observing runs and their run times:
\begin{equation}
    \alpha = \frac{d_{\text{BNS,O4}}^3T_{\text{O4}}}{d_{\text{BNS,O1}}^3T_{\text{O1}}+d_{\text{BNS,O2}}^3T_{\text{O2}}+d_{\text{BNS,O3}}^3T_{\text{O3}}}
\end{equation}
where $d_{\text{BNS},r}$ is the horizon distance of the detectors for a $(1.4\,M_{\odot},1.4\,M_{\odot})$ optimally oriented BNS corresponding to an SNR threshold of 8 given the sensitivity of the $r$th observing run and $T_r$ is the run-time of the $r$th observing run. With $d_{\text{BNS},r}$ and $T_r$ chosen from~\cite{O4projection} we are left with $\alpha=2.34$. Then choosing $n_0=2$ and $\beta=0$, we find the expected number of events to be $n=6^{+10}_{-5}$. These bounds on $n$ are computed by drawing samples from the distribution in Eq.~\eqref{p(n)} and computing the equal tail 90\% confidence intervals of those samples. Thus we draw 16 simulated events from the Galactic BNS population mentioned above for our simulation study so that the results of the study can also serve as a forecast of optimistic EoS  constraints from GW data one can expect to achieve at the end of O4.

\bibliographystyle{yahapj}
\bibliography{reference}
\end{document}